\documentclass[useAMS,usenatbib]{mn2e}

\usepackage{epsfig}
\usepackage{amsfonts}
\usepackage[usenames]{color}
\usepackage[usenames,dvipsnames]{xcolor}
\usepackage{url}
\usepackage{subfigure}
\usepackage{times,graphicx,amsmath,amsfonts,amssymb,aas_macros,epstopdf,hyperref}
\usepackage[normalem]{ulem}
\usepackage[T1]{fontenc}
\usepackage{multirow}
\usepackage{aecompl}

\usepackage{hyperref}
\hypersetup{
    colorlinks=true,
    linkcolor=blue,
    citecolor=blue,
}

\def\etal{{\frenchspacing\it et al.}}
\def\ie{{\frenchspacing\it i.e.}}
\def\eg{{\frenchspacing\it e.g.}}
\def\etc{{\frenchspacing\it etc.}}

\def\be{\begin{equation}}
\def\ee{\end{equation}}
\def\ba{\begin{eqnarray}}
\def\ea{\end{eqnarray}}

\def\mthod{$\mathrm{MTHOD}\,$}

\newcommand{\mpcoh}{\,h^{-1}\,{\rm Mpc}}

\def\d{{\rm d}}

\def\citejap#1{\citeauthor{#1}\ \citeyear{#1}}

\def\LaTeX{L\kern-.36em\raise.3ex\hbox{a}\kern-.15em
    T\kern-.1667em\lower.7ex\hbox{E}\kern-.125emX}

\title[A Multi-tracer analysis for BAO and RSD]{The clustering of the SDSS-IV extended Baryon Oscillation Spectroscopic Survey DR16 luminous red galaxy and emission line galaxy samples: cosmic distance and structure growth measurements using multiple tracers in configuration space}
\author[Wang \etal]{
\parbox{\textwidth}{
Yuting Wang$^{1}$\thanks{\url{Email: ytwang@nao.cas.cn}}, Gong-Bo Zhao$^{1,2}$\thanks{\url{Email: gbzhao@nao.cas.cn}}, Cheng Zhao$^{3}$, Oliver H.\,E. Philcox$^{4,5}$, Shadab Alam$^{6}$, Am\'elie Tamone$^{3}$, Arnaud de Mattia$^{7}$, Ashley J. Ross$^{8}$, Anand Raichoor$^{3}$, Etienne Burtin$^{7}$, Romain Paviot$^{9,10}$, Sylvain de la Torre$^{9}$, Will J. Percival$^{11,12,13}$, Kyle S. Dawson$^{14}$, H\'ector Gil-Mar\'{i}n$^{15,16}$, Julian E. Bautista$^{17}$, Jiamin Hou$^{18}$, Kazuya Koyama$^{17}$, John A. Peacock$^{6}$, Vanina Ruhlmann-Kleider$^{7}$, H{\'e}lion du Mas des Bourboux$^{14}$, Chia-Hsun Chuang$^{19}$, Johan Comparat$^{18}$, Stephanie Escoffier$^{10}$, Jean-Paul Kneib$^{3}$, Eva-Maria Mueller$^{20}$, Jeffrey A. Newman$^{21}$, Graziano Rossi$^{22}$, Arman Shafieloo$^{23}$, Donald P. Schneider$^{24}$} 
\vspace*{30pt} \\
$^{1}$ National Astronomy Observatories, Chinese Academy of Science, Beijing, 100101, P.R.China \\
$^{2}$ University of Chinese Academy of Sciences, Beijing 100049, P.R.China \\
$^{3}$ Institute of Physics, Laboratory of Astrophysics, \'Ecole Polytechnique F\'ed\'erale de Lausanne (EPFL), Observatoire de Sauverny, CH-1290 Versoix, Switzerland \\
$^{4}$ Department of Astrophysical Sciences, Princeton University, Princeton, NJ 08544, USA\\
$^{5}$ Harvard-Smithsonian Center for Astrophysics, 60 Garden St., MA, 02138, USA \\
$^{6}$ Institute for Astronomy, University of Edinburgh, Royal Observatory, Edinburgh, EH9 3HJ, United Kingdom \\
$^{7}$ IRFU, CEA, Universit\'e Paris-Saclay, F-91191 Gif-sur-Yvette, France \\
$^{8}$ Center for Cosmology and Astro-Particle Physics, Ohio State University, Columbus, Ohio, USA \\
$^{9}$ Aix Marseille Univ, CNRS, CNES, LAM, Marseille, France \\
$^{10}$ Aix Marseille Univ, CNRS/IN2P3, CPPM, Marseille, France \\
$^{11}$ Waterloo Centre for Astrophysics, University of Waterloo, Waterloo, ON N2L 3G1, Canada \\
$^{12}$ Department of Physics and Astronomy, University of Waterloo, Waterloo, ON N2L 3G1, Canada \\
$^{13}$ Perimeter Institute for Theoretical Physics, 31 Caroline St. North, Waterloo, ON N2L 2Y5, Canada \\
$^{14}$ Department Physics and Astronomy, University of Utah, 115 S 1400 E, Salt Lake City, UT 84112, USA \\
$^{15}$ Institut de Ci\`encies del Cosmos,   Universitat  de  Barcelona,  ICCUB,  Mart\'i  i  Franqu\`es  1,  E08028  Barcelona,  Spain \\
$^{16}$ Institut  d'Estudis  Espacials  de  Catalunya  (IEEC),  E08034  Barcelona,  Spain \\
$^{17}$ Institute of Cosmology \& Gravitation, University of Portsmouth, Dennis Sciama Building, Portsmouth, PO1 3FX, United Kingdom \\
$^{18}$ Max-Planck-Institut f\"ur Extraterrestrische Physik, Postfach 1312, Giessenbachstrasse 1, 85748 Garching bei M\"unchen, Germany \\
$^{19}$ Kavli Institute for Particle Astrophysics and Cosmology, Stanford University, 452 Lomita Mall, Stanford, CA 94305, USA \\ 
$^{20}$ Sub-department of Astrophysics, Department of Physics, University of Oxford, Denys Wilkinson Building, Keble Road, Oxford OX1 3RH\\
$^{21}$ PITT PACC, Department of Physics and Astronomy, University of Pittsburgh, Pittsburgh, PA 15260, USA \\
$^{22}$ Department of Physics and Astronomy, Sejong University, Seoul 143-747, Korea \\
$^{23}$ Korea Astronomy and Space Science Institute, 776 Daedeokdae-ro, Yuseong-gu, Daejeon 305-348, Republic of Korea \\
$^{24}$ Institute for Gravitation and the Cosmos, Pennsylvania State University, University Park, PA 16802, USA \\
\\
\\
\\
\\
\\
\\
\\
\\
\\
\\
\\
\\
}

\date{Accepted XXX. Received YYY; in original form \today}

\pubyear{2020}

\begin{document}
\label{firstpage}
\pagerange{\pageref{firstpage}--\pageref{lastpage}}
\maketitle

\begin{abstract} 
We perform a multi-tracer analysis using the complete Sloan Digital Sky Survey IV (SDSS-IV) extended Baryon Oscillation Spectroscopic Survey (eBOSS) DR16 luminous red galaxy (LRG) and the DR16 emission line galaxy (ELG) samples in the configuration space, and successfully detect a cross correlation between the two samples, and find the growth rate to be $f\sigma_8=0.342 \pm 0.085$ ($\sim25$ per cent accuracy) from the cross sample alone. We perform a joint measurement of the baryonic acoustic oscillation (BAO) and redshift space distortion (RSD) parameters at a single effective redshift of $z_{\rm eff}= 0.77$, using the auto- and cross-correlation functions of the LRG and ELG samples, and find that the comoving angular diameter distance $D_M(z_{\rm eff})/r_d = 18.85\pm 0.38$, the Hubble distance $D_H(z_{\rm eff})/r_d = 19.64 \pm 0.57$, and $f\sigma_8(z_{\rm eff}) = 0.432 \pm 0.038$, which is consistent with a $\Lambda$CDM model at $68\%$ CL. Compared to the single-tracer analysis on the LRG sample, the Figure of Merit (FoM) of $\alpha_{\perp}, \alpha_{||}$ and $f\sigma_8$ is improved by a factor of $1.11$ in our multi-tracer analysis, and in particular, the statistical uncertainty of $f\sigma_8$ is reduced by $11.6 \%$.
\end{abstract}

\begin{keywords} 
large scale structure of the Universe; cosmological parameters
\end{keywords}

\section{Introduction}
\label{sec:intro}
Observations of the large-scale structure of the Universe provide an essential probe of the physics of the accelerating cosmic expansion, which was discovered by the observation of type Ia supernovae \citep{Riess, Perlmutter}. The clustering analysis of large-scale structure allows us to measure the cosmic expansion history and structure growth via signals of baryon acoustic oscillations (BAO) and redshift space distortions (RSD), respectively \citep{Cole1995, Peacock2001, Cole2005, Hawkins2003, Eisenstein2005, Okumura2008, Percival2009}. The BAO, produced by the competition between gravity and radiation due to the coupling between baryons and photons before the cosmic recombination, leaves an imprint on the distribution of galaxies at late times. After the decoupling of photons, the acoustic oscillations are frozen at a characteristic scale around $\sim 100\,h^{-1}\rm Mpc$, which is determined by the comoving sound horizon at the drag epoch $r_d$. This feature corresponds to an excess in the 2-point correlation function, or a series of wiggles in the power spectrum \citep{Percival:2001hw, Cole2005, Eisenstein2005}, making BAO a robust observable as a cosmic standard ruler. Measuring the BAO scale in the radial and transverse directions provides strong constraints on Hubble expansion rate and angular diameter distance, respectively. The RSD is produced due to peculiar motions of galaxies: galaxies tend to infall towards the local over-density regions, thus the clustering along the line-of-sight (LOS) is enhanced on large scales \citep{Kaiser1987, Peacock2001}.Thus measuring RSD effect sets a constraint on the growth rate of cosmic structure. 

The most precise BAO and RSD measurements to date were reported by the Baryon Oscillation Spectroscopic Survey (BOSS) collaboration using the final Data Release 12 (DR12) \citep{Alam}, which contains more than one million galaxies with spectroscopic redshifts. BOSS achieved a $\smash{(1.0 -2.5)}$ per cent BAO measurement precision and a $9.2$ per cent RSD precision in the redshift range of $0.2<z<0.75$ \citep{Alam2016}, and extracted tomographic information of galaxy clustering in the past lightcone \citep{Wang:2016wjr,Zhao:2016das,Wang:2017wia,Zheng:2018kgq}, which is key for probing dynamical dark energy \citep{Zhao:2017cud,Wang:2018fng}. The BOSS DR12 data can provide high-precision constraints on cosmological parameters \citep{Ivanov:2019pdj,DAmico:2019fhj,Colas:2019ret,Philcox:2020vvt}. The extended BOSS (eBOSS) project, the sussessor of BOSS, aims to map the Universe using multiple galaxies at higher redshifts, covering the redshift range of $0.6<z<2.2$ \citep{Dawson2016}. It allows for BAO and RSD measurements at high redshifts, which is crucial to break degeneracy between key cosmological parameters, \eg\, $H_0$ and $\Omega_m$ \citep{Wang:2017yfu}.

However, the precision of the measurements of galaxy clustering is restricted by the cosmic variance on large scales due to the limited volume that a galaxy survey can map, and by the shot noise on small scales due to the discreteness of galaxies. One potential way to tackle the cosmic variance is to contrast multiple tracers of the dark matter field with different biases, \ie\, the `multi-tracer' technique \citep{McDonald:2008sh, Seljak:2008xr}. In the ideal case with no shot noise, the ratio of over-densities of two tracers would be independent of the density field of dark matter, then the measurements of parameters related to the bias parameter can be immune to the cosmic variance, and thus they can be accurately determined. For practical applications, the gain from multiple tracers can be downgraded by various factors including the overlapping redshift ranges and sky regions, the ratio of biases, the Poisson noise of the 2-point function of each tracer, \etc \,Multi-tracer studies of galaxy surveys have been performed; for instance, \citet{Blake:2013nif} found a $10-20$ per cent improvement on the RSD measurement via the multi-tracer analysis of the Galaxy and Mass Assembly survey (GAMA). This technique was also applied to analysing the galaxy clustering in the overlapping region between the BOSS and WiggleZ surveys \citep{Ross:2013vla, Beutler:2015tla, Marin:2015ula}. 

The eBOSS survey, which is a part of the Sloan Digital Sky Survey-IV (SDSS-IV) project \citep{sdss4tech}, used the 2.5-metre Sloan telescope \citep{sdsstelescope} located at the Apache Point Observatory in New Mexico. The spectra of samples are collected by the two multi-object fiber spectrographs \citep{sdssspectrograph}. eBOSS is the first survey that can simultaneously observe multiple galaxies with large overlapping areas in a broad redshift range, which is ideal for a multi-tracer analysis. In this paper we present a multi-tracer analysis using the final eBOSS DR16 Luminous Red Galaxy (LRG) sample combined with the high redshift tail from BOSS DR12 CMASS (for ``Constant stellar Mass") sample, dubbed `LRGpCMASS' sample, and the eBOSS DR16 Emission Line Galaxy (ELG) sample. 

This work is one of a series of papers presenting results based on the final eBOSS DR16 samples. The multi-tracer analysis of the same samples is also performed in Fourier space to complement this work \citep{ZhaoGB2020}. For the LRG sample, produced by \citet{eBOSSDR16catalogue}, the correlation function is used to measure BAO and RSD in \citet{Bautista2020}, and the analyses of BAO and RSD from power spectrum are discussed in \citet{GilMarin2020}. The LRG mock challenge for assessing the modelling systematics is described in \citet{Rossi2020}. The ELG catalogues are presented in \citet{ELGcatalogue}, and analyzed in Fourier space \citep{deMattia2020} and in configuration space \citep{Tamone2020}, respectively. The clustering catalogue of quasars is generated by \citet{eBOSSDR16catalogue}. The quasar mock challenge for assessing modelling systematics is described in \citet{Smith2020}. The quasar clustering analysis in Fourier space is discussed in \citet{Neveux2020}, and in configuration space in \citet{Hou2020}. Finally, the cosmological implications from the clustering analyses is presented in \citet{Mueller2020}.

We introduce the galaxy samples and mock catalogues used in this paper in Sections \ref{sec:data} and \ref{sec:mock}, respectively. In Section \ref{sec:model}, we describe the template of the full shape correlation function, and in Section \ref{sec:measureCF}, we show measurements of the correlation function. The methodology of parameter estimation and the fitting result are presented in Sections \ref{sec:BAORSDfit}, \ref{sec:mockresult}, and \ref{sec:dataresult} respectively. We discuss cosmological implications using in Section \ref{sec:application}. Section \ref{sec:conclusion} is devoted to the conclusion. In this paper, we use a fiducial $\Lambda$CDM cosmology with parameters:  $\Omega_m=0.307, \Omega_bh^2=0.022, h=0.6777, n_s=0.96,  \sigma_8=0.8288$. The comoving sound horizon in this cosmology is $r_d^{\rm fid}=147.74 \,\rm Mpc$.

\section{Galaxy samples} 
\label{sec:data}
In this section, we briefly describe the eBOSS DR16 galaxy sample used in the work.

\begin{figure}   
\centering
\includegraphics[scale=0.35]{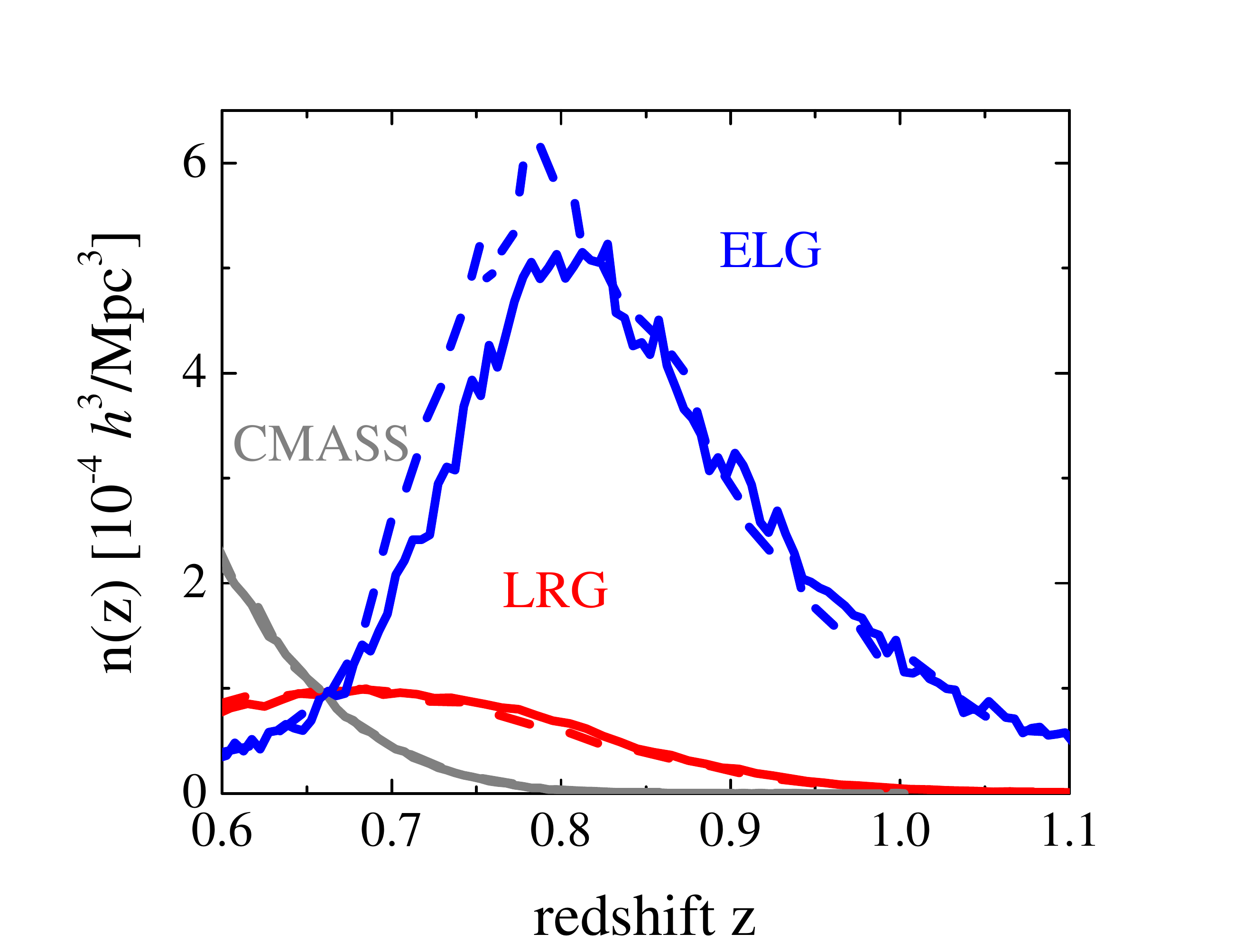}
\caption{The volume density as a function of redshift for eBOSS DR16 LRG (red), BOSS DR12 CMASS (grey), and eBOSS DR16 ELG (blue) samples. The distribution in North Galactic Cap is shown in solid curves and South Galactic Cap in dashed curves.}\label{fig:nz}
\end{figure} 

\subsection{The eBOSS LRG and BOSS CMASS samples}
\begin{figure*}   
\includegraphics[scale=0.2]{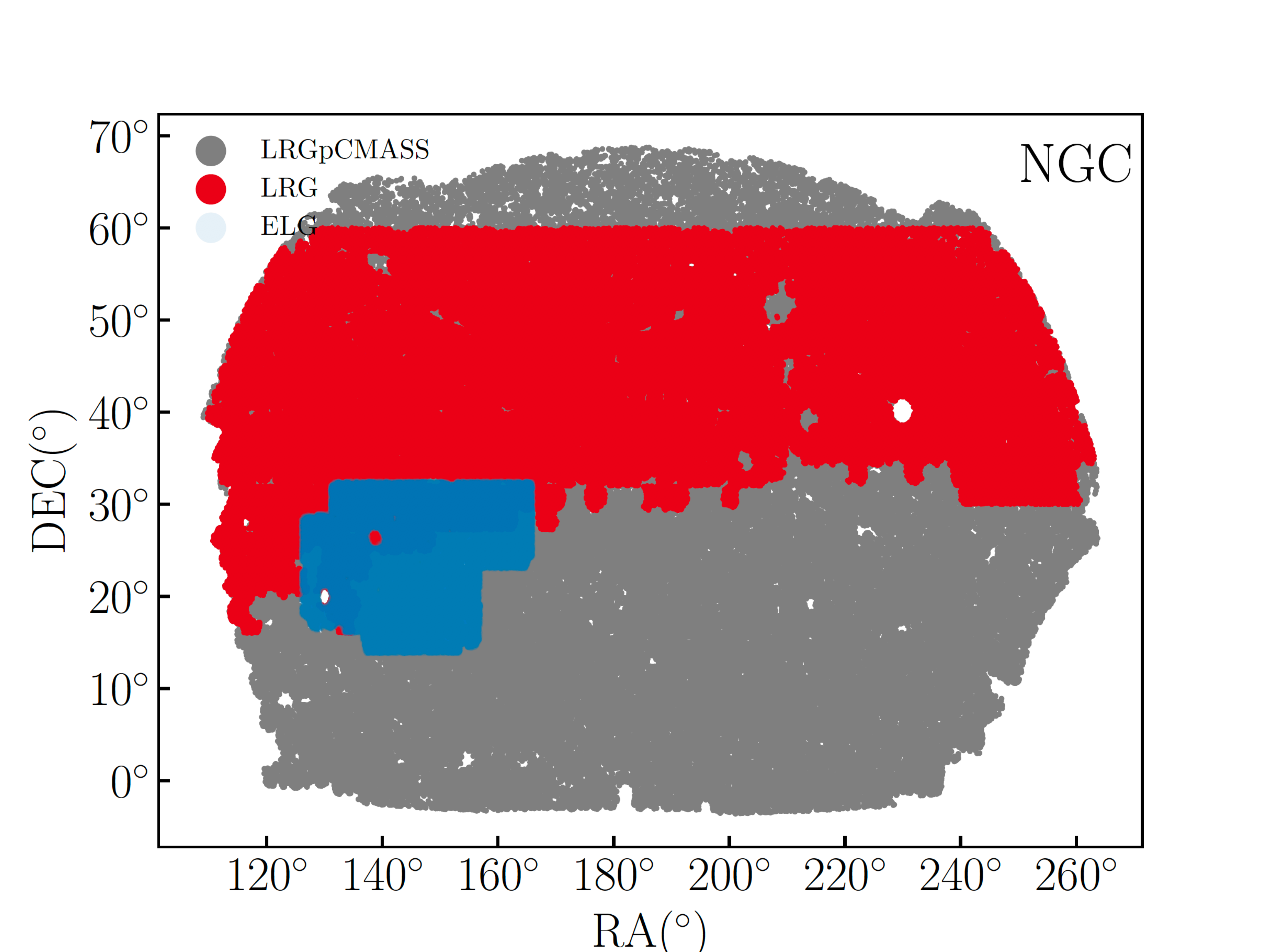}
\includegraphics[scale=0.2]{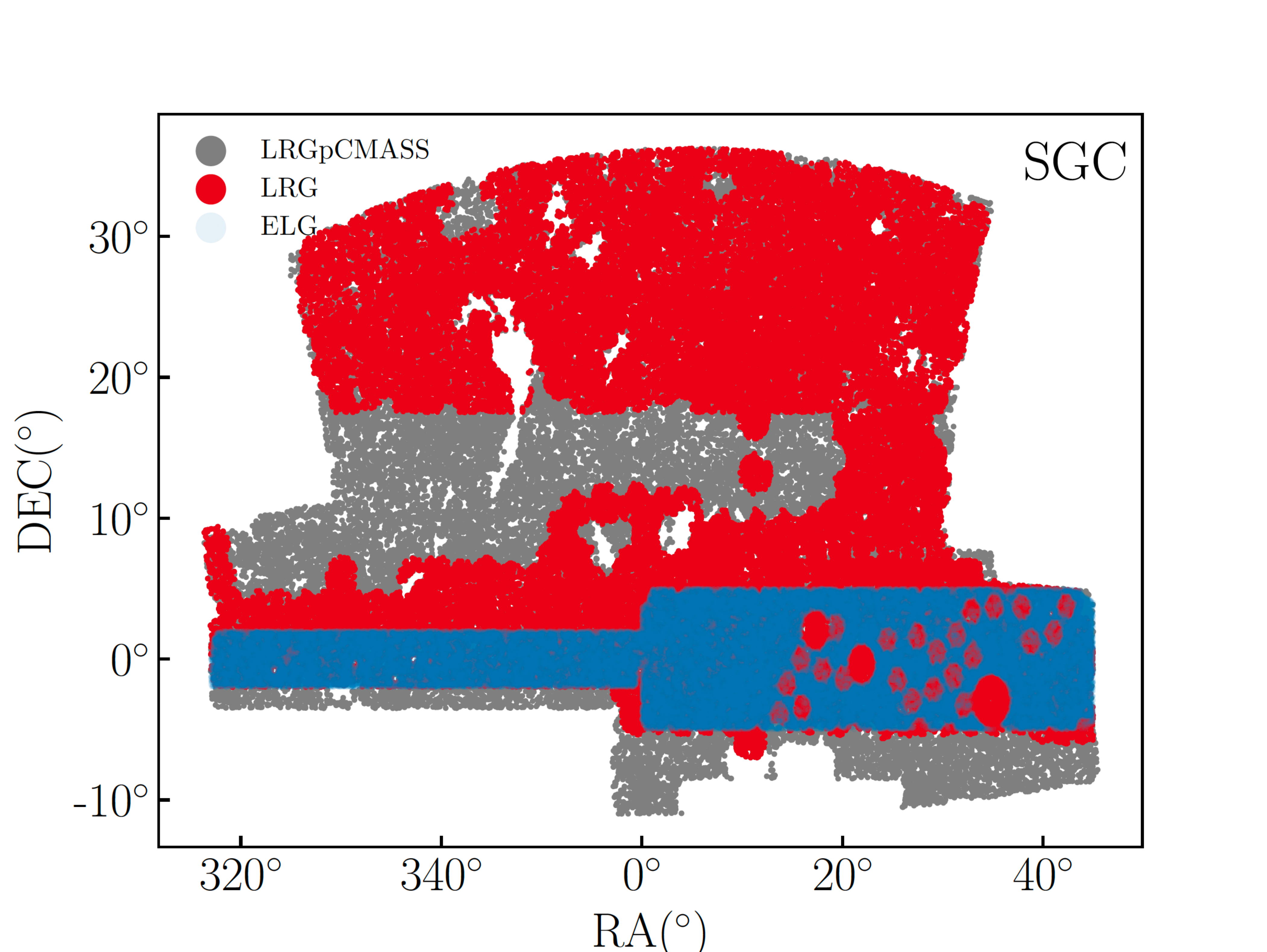}
\caption{Footprint of eBOSS DR16 LRG (red) and ELG (blue), and a combined sample of eBOSS DR16 LRG and BOSS DR12 CMASS (grey) in the North Galactic Cap (NGC, left) and South Galactic Cap (SGC, right).}\label{fig:sky}
\end{figure*} 
The target sample of luminous red galaxies was selected from the optical SDSS photometry DR13 \citep{Albaretietal13} and the infrared photometry from the WISE satellite \citep{Langetal14}. The final algorithms for target selection and catalogue generation are described in \cite{Prakashetal16} and in a companion paper \citep{eBOSSDR16catalogue}. We use the LRG data of the complete $5$ years of eBOSS in the redshift range of $0.6 < z< 1.0$. Its volume density distribution as a function of redshift is shown in red curves in Figure \ref{fig:nz}. The sky coverage of eBOSS DR16 LRG is $2475.51\,\rm deg^2$ in the North Galactic Cap (NGC) and $1626.80\,\rm deg^2$ in the South Galactic Cap (SGC), which are shown in red regions of Figure \ref{fig:sky}. 

In order to correct for observational effects, the eBOSS DR16 LRG catalogue is assigned a set of weights, including weights for the redshift failure, $w_{\rm zf}$, close pair due to fiber collisions, $w_{\rm cp}$ and for systematics due to the effect of completeness, the variation of the mean density as a function of stellar density and Galactic extinction, $w_{\rm sys}$. In addition, the FKP weight to minimize the variance in the clustering measurement combining regions \citep{FKP1994} is added 
\begin{equation}
w_{\rm FKP}=\frac{1}{1+n(z)P_0}\,,
\end{equation}
where $n(z)$ is the number density of galaxies, and $P_0$ is set to $10,000\,h^{-3}\rm Mpc^3$. The total weight applied to each eBOSS LRG is \citep{eBOSSDR16catalogue} 
\begin{equation}
w^{\rm LRG}_{\rm tot}=w_{\rm FKP}\times w_{\rm sys}\times w_{\rm cp} \times w_{\rm zf}\,.
\end{equation}

The eBOSS DR16 LRG sample overlaps with the BOSS DR12 CMASS in the redshift range of $0.6<z<1.0$ \citep{Reid2016}, as shown in Figure \ref{fig:sky}, thus these two catalogs are merged into one sample, dubbed `LRGpCMASS', in this work. Note that the BOSS DR12 CMASS used different procedures for generating close-pair and redshift failure weights and the total weight is counted via \citep{Reid2016} 
\begin{equation}
w^{\rm CMASS}_{\rm tot}=w_{\rm FKP}\times w_{\rm sys}\times \left(w_{\rm cp} + w_{\rm zf}-1\right)\,.
\end{equation}
The combined DR16 BOSS+eBOSS LRG catalogue includes the correct total weight for each LRG in order to avoid confusion (see Section 5.7 of \citealt{eBOSSDR16catalogue} for more details).

\subsection{The eBOSS ELG sample}
\label{data:ELG}
The target sample of emission line galaxies is selected from the DECam Legacy Survey (DECaLS) $grz-$photometry \citep{raichoor17}, which provides an imaging dataset at higher redshifts. The final large-scale structure catalogue creation is described in the companion paper \citep{ELGcatalogue}. We use the complete eBOSS DR16 ELG catalogues in the redshift range of $0.6 <z <1.1$, which is composed of $83,769$ galaxies in the NGC and $89,967$ galaxies in the SGC with spectroscopic redshifts. The redshift distributions in NGC and SGC are shown in blue solid and dashed curves in Figure \ref{fig:nz}. The eBOSS DR16 ELG sample overlaps with LRGpCMASS within $0.6 <z <1.0$. The effective sky area of ELG is $369.4 \rm \, deg^2$ in NGC and $357.5 \rm \, deg^2$ in SGC, which are shown in blue regions of Figure \ref{fig:sky}. The overlapping area covered ELG and LRGpCMASS samples is $\sim 730\, \rm deg^2$.  

The total weight assigned to each eBOSS ELG is
\begin{equation}
w^{\rm ELG}_{\rm tot}=w_{\rm FKP}\times w_{\rm sys}\times w_{\rm cp} \times w_{\rm zf}\,;
\end{equation}
here $P_0=4000\,h^{-3}\rm Mpc^3$ in $w_{\rm FKP}$. A description of the observational effects is presented in \citet{ELGcatalogue}.

The ELG sample suffers from angular systematics, which could be due to the photometry of the imaging observation used for target selection, and this kind of observation systematics may bias the measurement of galaxy clustering \citep{deMattia2020,Tamone2020}. \citet{Burden2017} proposed a modified model of correlation function to null the angular modes from the galaxy clustering, such that the contamination from angular systematics can be largely reduced. A sophisticated model is developed by \citet{Paviot2020}, which is used for this analysis.

\subsection{The radial integral constraint}
\label{sec:ric}
The true radial selection function in spectroscopic surveys is difficult to determine from the survey itself, and it is commonly approximated from the redshift distribution of the actual data sample. When generating the corresponding random catalogue, the redshifts of data are assigned to the random catalogues, dubbed the $\mathit{shuffled}$ scheme. This ensures that the average density fluctuations along the LOS are zero, but leads to an impact on the galaxy clustering on large scales. This effect is called as the radial integral constraint (RIC). The scheme to correct the RIC effect in theory was proposed by \citet{de-Mattia:2019aa}. This modelling method is used to account for the correction of RIC effect in the analysis of eBOSS DR16 ELG clustering (see \eg\,~\cite{deMattia2020, Tamone2020}). Alternatively, we can subtract the RIC effect from the data measurement. Firstly we quantify the RIC effect using additional two sets of EZmocks without systematics \citep{ZhaoC2020}. One set of mocks contains the RIC effect, in which the redshifts of the random catalogues are assigned from the redshifts of each mock data via the $\mathit{shuffled}$ scheme. The other set is without the RIC effect, where $1000$ mock datasets use a single random catalogue sampling the redshift distribution of data (dubbed the $\mathit{sampled}$ scheme). The difference between these two sets of mocks provides an estimation of the RIC effect, which then can be subtracted from the data measurement. We are aware that this is an approximation, as the dependence of the RIC on cosmological parameters is not accounted for in this scheme. We performed a comparison with the forwarding modelling method and find the difference is negligible given the statistical uncertainty of the ELG sample. 

\subsection{The effective redshift}
The effective redshift of the sample is determined via the following weighted pair-count,
\begin{equation}
z_{\rm eff}=\frac{\sum w^m_iw^n_j(z^m_i+z^n_j)/2}{\sum w^m_i w^n_j}\,,
\label{eq:zeff}
\end{equation}
where $w_i$ is the total weight of the $i$th galaxy at redshift $z_i$. We compute the effective redshift over all galaxy pairs separated by a distance between $25$ and $150\,h^{-1}{\rm Mpc}$,\footnote{The limits of separations have little effect on the value of the effective redshift.} having $z_{\rm eff}=0.70$ for the combined sample of NGC and SGC LRGpCMASS when $m=n={\rm L}$, $z_{\rm eff}=0.845$ for the ELG combined sample in NGC and SGC when $m=n={\rm E}$, and $z_{\rm eff}=0.77$ for the cross galaxy pairs between LRGpCMASS and ELG samples, \ie\, $m={\rm L}$ and $n={\rm E}$.

\section{Mock catalogues}
\label{sec:mock}

In this section we present the mock datasets, on which we will perform series of tests to check our pipeline of analysis, including the modeling and parameter estimation.

\subsection{MDPL2 mocks}
To test our modelling of non-linear gravitational collapse and certain aspects of galaxy physics, we generate mock catalogues using the Multi-tracer Halo Occupation Distribution \citep[\mthod;][]{Alam:2019pwr}.
The \mthod approach introduces a new way to model multiple tracers in the same volume. In this approach each of the tracers can have its own occupation recipe for the central and satellite galaxies. \mthod ensures that the joint probabilities of occupation are well behaved by limiting the total probability of central galaxies in a halo to $1$ and makes sure that non-physical behaviour is forbidden, such as multiple types of galaxies at the centre of the same dark matter halo. The key parameters in \mthod models involve the independent parameters for the occupation probability of central and satellite galaxies for each tracer.
The \mthod mock galaxy catalogue is created using the MultiDark Planck simulation \citep[MPDL2; ][]{2012MNRAS.423.3018P} publicly available\footnote{\url{https://www.cosmosim.org/cms/simulations/mdpl2/}} through the CosmoSim database. MPDL2 is a dark matter only $N$-body simulation using the Gadget-2 algorithm \citep{2016MNRAS.457.4340K}.
MDPL2 assumes a flat $\Lambda$CDM cosmology with $\Omega_m=0.307$, $\Omega_b=0.048$, $h=0.67$, $n_s=0.96$ and $\sigma_8=0.82$, and is a periodic box of side length 1$\mathrm{h^{-1}Gpc}$ sampled by $3840^3$ particles. A halo catalogue is generated using the ROCKSTAR halo finder \citep{behroozi13} at an effective redshift of $z = 0.86$. 

The DM haloes are then populated using the following equations for central and satellite galaxies as a function of halo mass, $M_{\rm halo}$:
\begin{align}
     p_{\rm cen}^{\rm tot}(M_{\rm halo};\vec{\theta}) &=\sum_{{\rm tr} \in {\rm TR}} p_{\rm cen}^{\rm tr}(M_{\rm halo};\theta^{\rm tr})\\
      \left< N_{\rm sat}^{\rm tot} \right>(M_{\rm halo}; \vec{\theta}) &=\sum_{{\rm tr} \in {\rm TR}} \left< N_{\rm sat}^{\rm tr} \right>(M_{\rm halo};\theta^{\rm tr}),
\end{align}
where the sum is over all tracers in the list, ${\rm TR} = \left\{ \rm LRG,QSO,ELG \right\}$. This equation requires a constraint of $p_{\rm cen}^{\rm tot}\leq1$ for any halo mass. 
The explicit forms of $p_{\rm cen}^{\rm tr}$ and $\left< N_{\rm sat}^{\rm tr} \right>$ are given in equations 8-14 in \citet{Alam:2019pwr}.  The full list of parameters ($\vec{\theta}=\left\{ \theta^{\rm LRG},\theta^{\rm ELG},\theta^{\rm QSO}\right\}$) and best-fit values obtained for the eBOSS samples are given table 1 of \citet{Alam:2019pwr} . 
All three tracers (i.e. LRG, ELG and QSOs) are modelled within the \mthod framework. However, we only use the LRG and ELG galaxies, and do not use the QSOs from the default in this paper. The number of LRG galaxies is $156,800$ and the number of ELG galaxies is $3,301,753$, with a much higher volume density of ELGs than that of LRGs. Two different models are used to populate the central galaxy called standard HOD (SHOD) and High Mass Quenched (HMQ) model. We create six realisations for each mock catalogue by projecting RSD along different axes of the cubic box.

\subsubsection{The semi-analytic covariance matrix}
In this analysis, we have six non-trivial combinations of correlation function multipoles, each of which has $25$ bins. This leads to a total of $11325$ independent covariance matrix elements, thus the covariance requires significant computational power to compute. As an alternative, we consider semi-analytic methods, in particular the \texttt{RascalC} method \citep{2019MNRAS.490.5931P,2020MNRAS.491.3290P}, which is a fast algorithm for computing two- and three-point correlation function covariances in arbitrary survey geometries. This works by noting that, in the Gaussian limit, the covariance can be written as an integral of products of the correlation function over four copies of the survey window function, which can be rapidly evaluated using importance sampling and random particle catalogues. Non-Gaussianity can be added via a small rescaling of the shot-noise terms, shown to be an excellent approximation on BAO scales in \citet{2016MNRAS.462.2681O} and \citet{2019MNRAS.487.2701O}). Using \texttt{RascalC} it is possible to estimate covariance matrices from an observational dataset and window function alone, drastically reducing the dependence on mocks and hence the computational resources required.

Here, we estimate the covariances for the periodic MDPL2 mocks, using all non-trivial combinations of LRG, ELG and cross correlation functions. As an input we require estimates of the correlation function computed over a large range of radii; these are estimated from the mocks using bins of width $\Delta r = 2\mpcoh$ from $r = 0\mpcoh$ to $r = 200\mpcoh$ and ten angular bins. For efficient configuration-space sampling we use random particle catalogues, which, given the periodic geometry, are here simply sets of $\sim 10^6$ particles uniformly placed on the cube for both LRGpCMASS and ELG samples. In total, we sample $\sim 10^{14}$ quadruplets of points in configuration-space to build a smooth model, which requires $\sim 400$ CPU-hours in total, significantly less than that required for traditional mock-based analyses.

\subsection{The EZmocks}
To estimate the covariance matrices of the clustering measurements of the full eBOSS data, we rely on $1000$ realisations of multi-tracer EZmock catalogues, for both LRGs and ELGs. These mocks are based on dark matter density fields generated using the Zel'dovich approximation \citep{Zeldovich1970}. Galaxies are then sampled in the density field with effective bias descriptions. The bias models for LRGs and ELGs are calibrated separately to the eBOSS data, with four free parameters. Nevertheless, the underlying dark matter density fields for different tracers are evolved from the same initial conditions, to account for their cross correlations. As the result, the cross correlation function between the EZmock LRGs and ELGs are well consistent with that of the data on small scales \citep[for details, see][]{ZhaoC2020}. 

In this work we use three different sets of EZmocks. Two of them are free of observational systematics, with only survey footprint, veto masks, and radial selections applied, which are used to estimate the RIC effect mentioned in Sec.\,\ref{sec:ric}. The random catalogues for these two sets of mocks are generated using the {\it sampled} and {\it shuffled} schemes respectively. For the {\it sampled} random catalogues, the redshift distributions are sampled from the spline-smoothed $n(z)$ of the data, while for the {\it shuffled} randoms, the redshifts are taken directly from the corresponding galaxy catalogues. The third set of EZmocks contain various observational effects, such as photometric systematics, fiber collisions, and redshift failures. These contaminated mocks are used to measure the covariance matrices of our analysis.

\begin{figure}   
\centering
\includegraphics[scale=0.55]{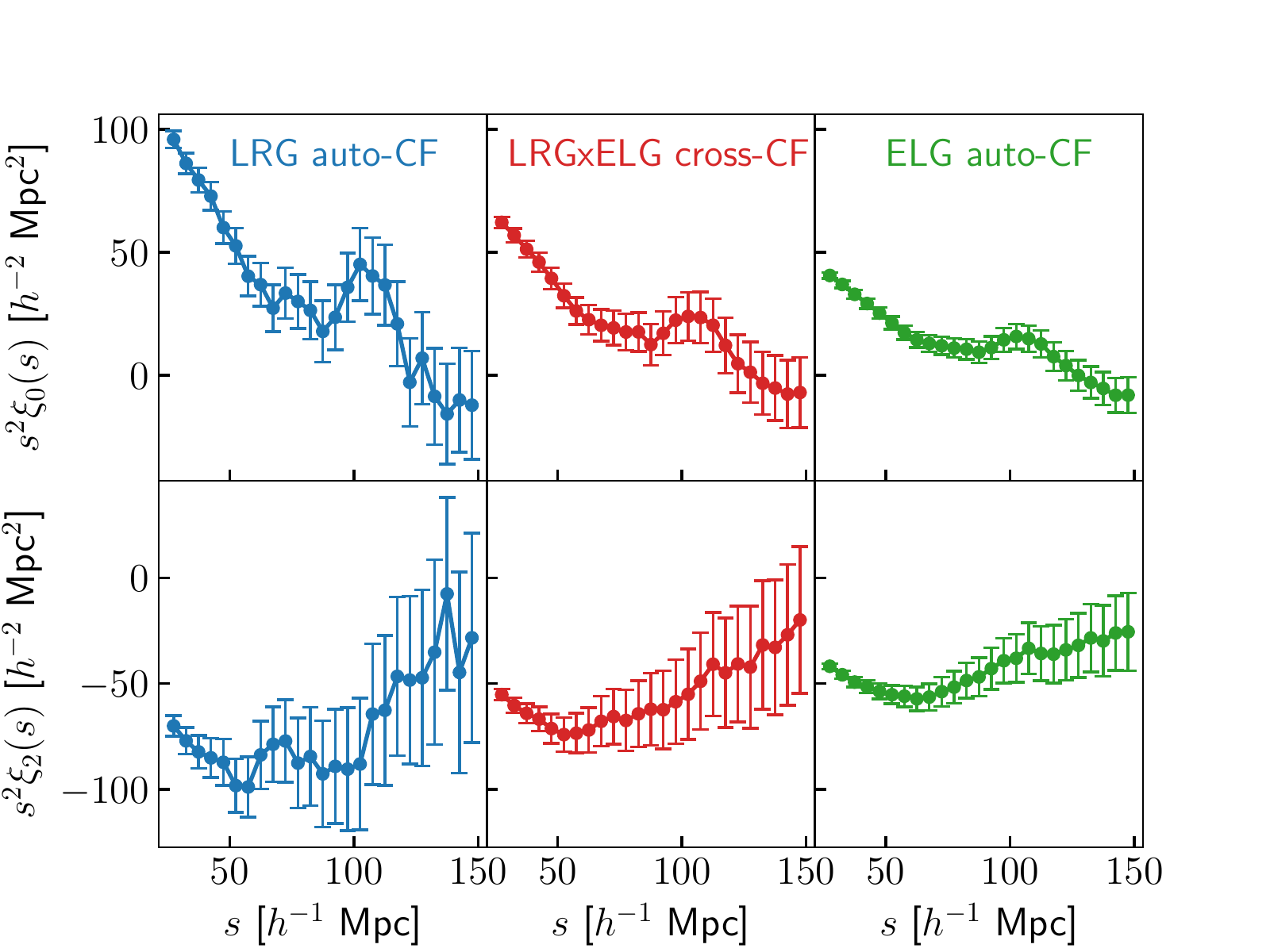}
\caption{The measured monopole (upper panels) and quadrupole (bottom panels) of the correlation function from a set of MDPL2 mocks following the multi-tracer HMQ HOD model. The LOS is set to be along the $z$ axis. The $1\,\sigma$ error bar is estimated from the RascalC covariance matrix.}\label{fig:xilMockCh}
\end{figure}

\begin{figure*}   
\includegraphics[scale=0.5]{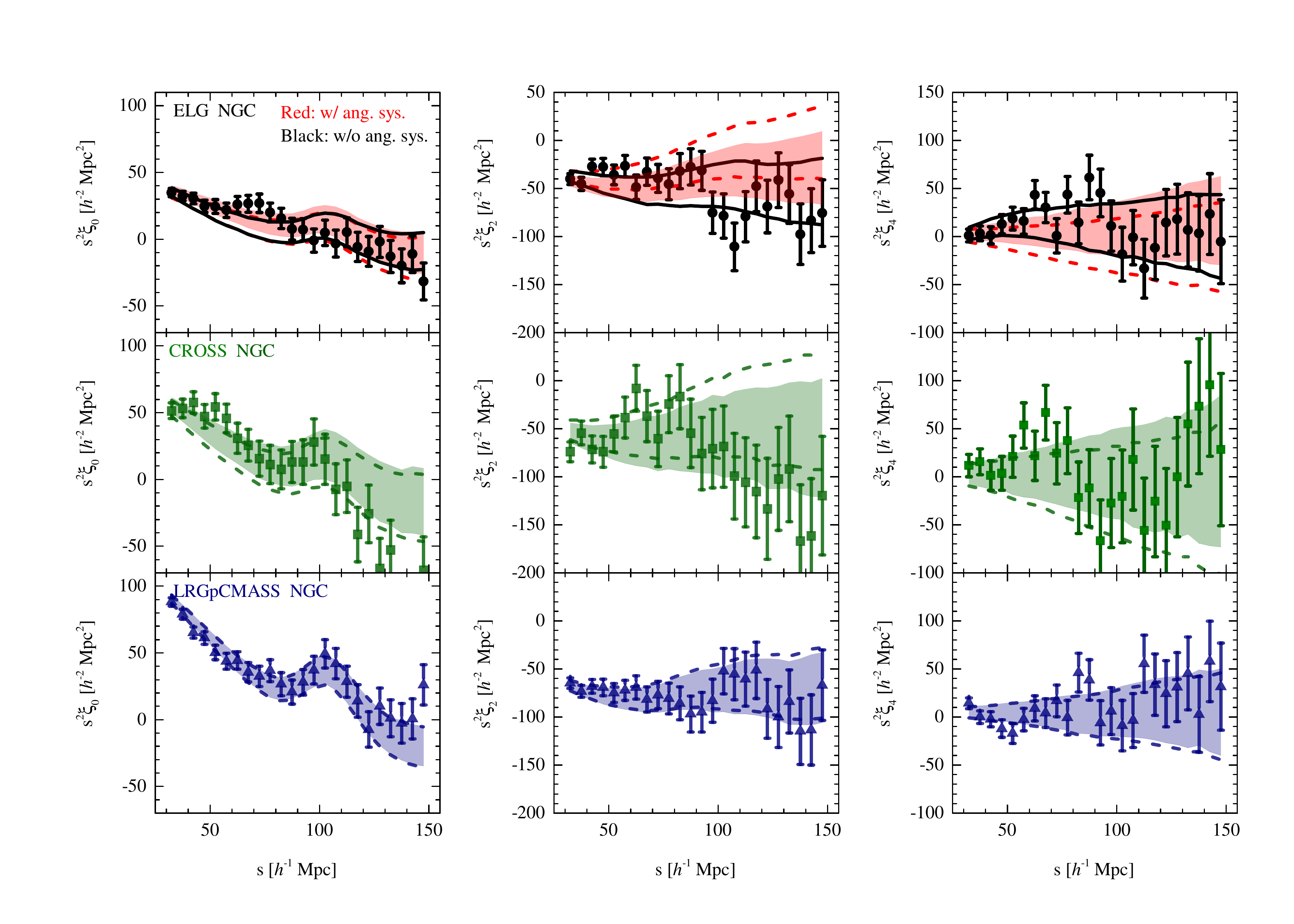}
\caption{The measured monopole, quadrupole, and hexadecapole of correlation functions for the ELG (top, red) and LRGpCMASS (bottom, blue) samples, and their cross-correlation (middle, green) in the NGC. The $1\,\sigma$ error bar is calculated from EZmock covariance matrix. The dashed areas and shaded bands in each panel are the averages of multipoles with a standard deviation from EZmocks with RIC and without RIC, respectively. For the ELG sample (top panels), the black solid circles (the measurements of data samples with $1\,\sigma$ error bars) and black-lines regions (the mean of $1000$ EZmock measurements with a standard deviation) are the measurements with the angular systematics corrected.}\label{fig:xilNGC}
\end{figure*} 

\begin{figure*}   
\includegraphics[scale=0.5]{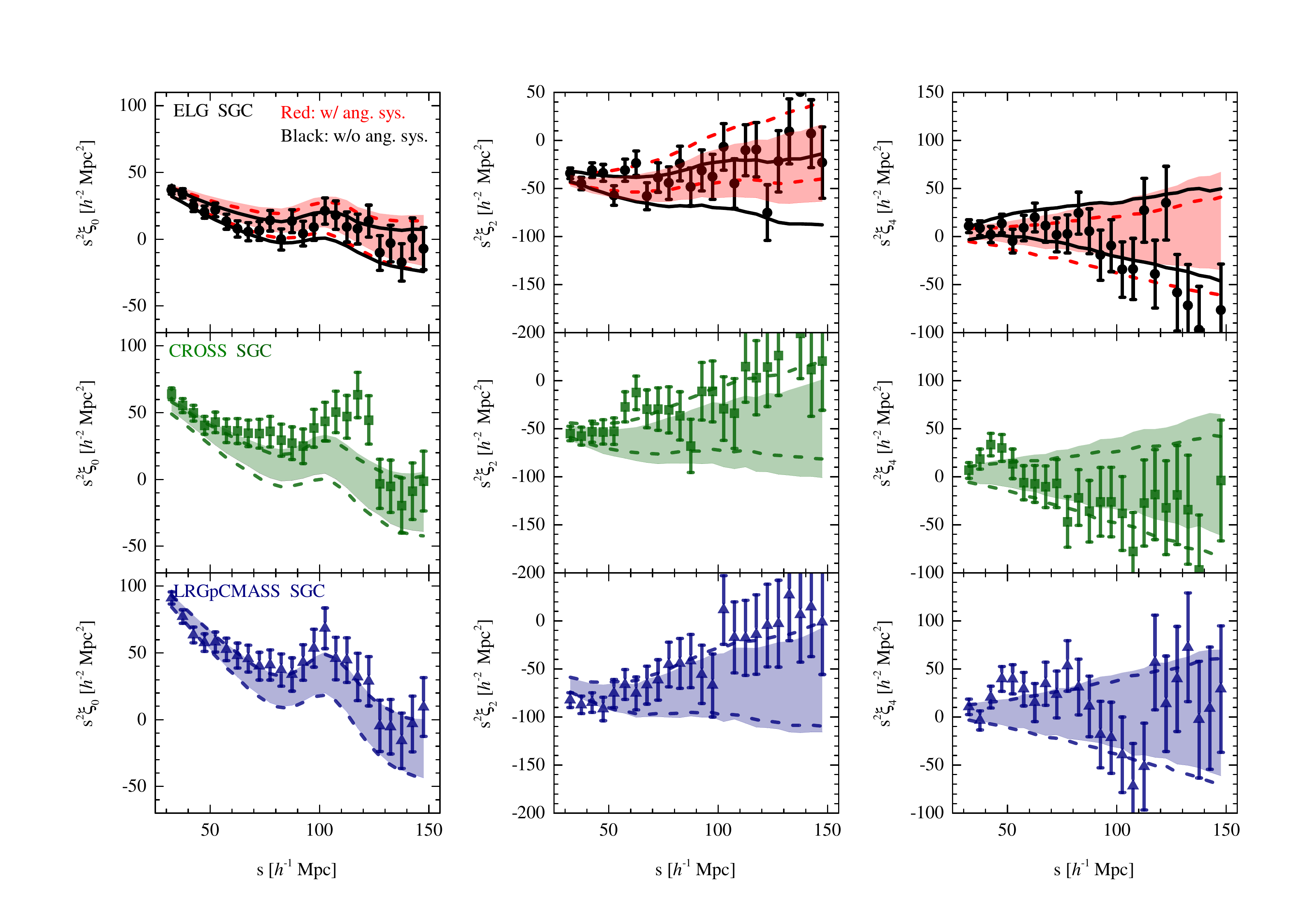}
\caption{As Figure \ref{fig:xilNGC}, but for the SGC.}\label{fig:xilSGC}
\end{figure*} 

\section{The template for the full shape analysis} 
\label{sec:model}
We use the `Gaussian streaming model' (GSM) developed in \citet{Reid2011} to compute the theoretical correlation function:
\begin{eqnarray}
1+\xi(s_{\perp}, s_{\parallel})&=&\int \frac{\d y }{\sqrt{2\pi \left[\sigma^2_{12}(r, \mu)+\sigma^2_{\rm FoG}\right]}}  \left[1+\xi(r)\right] \nonumber \\
&\times& \exp \left\{-\frac{\left[s_{\parallel} - y - \mu v_{12}(r)\right]^2}{2 \left[\sigma^2_{12}(r, \mu)+\sigma^2_{\rm FoG}\right] }\right\},
\label{eq:streaming} 
\end{eqnarray}
where $\smash{s_{||} \equiv s \mu}$ and $\smash{s_{\perp}\equiv s \sqrt{(1-\mu^2)}}$ 
denotes the separation of pairs along and across the LOS, respectively; $\xi(r)$ is the real-space correlation function as a function of the real-space separation $r$; $v_{12}(r)$ is the mean infall velocity of galaxies separated by $r$; and $\sigma_{12}(r, \mu)$ is the pairwise velocity dispersion of galaxies. The parameter $\sigma_{\rm FOG}$ is used to marginalize over the Fingers-of-God (FoG) effect on nonlinear scales due to random motions of galaxies. The quantities $\xi(r)$, $v_{12}(r)$ and $\sigma_{12}(r, \mu)$ are computed using the Convolution Lagrangian Perturbation Theory (CLPT),\footnote{\url{https://github.com/wll745881210/CLPT_GSRSD}} \citep{Carlson:2012bu, Wang:2013hwa} 
\begin{eqnarray}
1 + \xi(r) &=& \displaystyle \int d^3q M_0(r,q) \,\\
v_{12,i}(r) &=& \left[1 + \xi(r)\right]^{-1} \displaystyle \int d^3q M_{1,i}(r,q)\,\\
\sigma_{12,ij}^2(r) &=& \left[1 + \xi(r)\right]^{-1} \displaystyle \int d^3q M_{2,ij}(r,q)\,,
\end{eqnarray}
where $v_{12,i}(r)$ is the component of mean pairwise velocity along the direction of pairwise separation $\hat{r}_i$, and $\sigma_{12,ij}^2(r)$ is the velocity dispersion component along the pairwise separation vector $\hat{r}$. $M_0(r,q)$, $M_{1,i}(r,q)$ and $M_{2,ij}(r,q)$ are the  convolution kernels that depend on the linear matter power spectrum and the first two non-local derivatives of the Lagrangian bias, $i.e.$ $\left\langle F' \right\rangle$ and $\left\langle F'' \right\rangle$ (see \citejap{Wang:2013hwa} for more details).

As mentioned in Sec.\,\ref{data:ELG}, for the auto-correlation function of ELG, we need to account for a correction to the angular systematics in the modelling. Such a template of the modified correlation function, as shown below, developed by \citet{Paviot2020} can well mitigate the angular contamination.
\begin{eqnarray}
\label{eq:modifiedGSM}
\tilde{\xi}(s_{\perp},s_{||}) &=& \xi(s_{\perp},s_{||}) \nonumber \\ 
&-& \frac{2\int \xi(s_{\perp},s_{||}') n\left[\chi(z_{\rm RP})-s_{||}'/2\right]ds_{||}'}{ \int n(\chi)d\chi} \nonumber \\ 
&+& \frac{\int n^2(\chi) d\chi \int \xi(s_{\perp},s_{||}') ds_{||}'}{\left[\int n(\chi) d\chi \right]^2} \,,
\end{eqnarray}
where $n(\chi)$ is the radial selection function of the survey, $\chi$ is the comoving distance out to a galaxy at redshift $z$, and the parameter $z_{\rm RP}=0.84$ is determined by minimizing the difference between the mean of the modified correlation function multipoles from two sets of ELG EZmocks (with and without systematics), as performed in \cite{Tamone2020}. 

The CLPT-GSM model can be easily generalised to model the cross-correlations between two tracers with different biases via the following transformation \citep{Carlson:2012bu, Wang:2013hwa},
\begin{eqnarray}
\label{eq:biasCross}
\left\langle F' \right\rangle &\rightarrow& \frac{1}{2}\left(\left\langle F'_m \right\rangle + \left\langle F'_n \right\rangle \right) \\
\left\langle F'' \right\rangle &\rightarrow& \frac{1}{2}\left(\left\langle F''_m \right\rangle + \left\langle F''_n \right\rangle \right) \\
\left\langle F' \right\rangle ^2 &\rightarrow& \left\langle F'_m \right\rangle \left\langle F'_n \right\rangle  \\
\left\langle F'' \right\rangle ^2 &\rightarrow& \left\langle F''_m \right\rangle \left\langle F''_n \right\rangle  \\
\left\langle F' \right\rangle\left\langle F'' \right\rangle &\rightarrow& \frac{1}{2}\left(\left\langle F'_m \right\rangle \left\langle F''_n \right\rangle + \left\langle F''_m \right\rangle \left\langle F'_n \right\rangle\right) \,.
\end{eqnarray}
Here the first local Lagrangian bias $\left\langle F' \right\rangle$ is related to the Eulerian linear bias factor $b$ via,
\begin{equation}
b = 1 + \left\langle F' \right\rangle \,,
\end{equation}
and the second local Lagrangian bias $\left\langle F'' \right\rangle$ is fixed under the peak-background split assumption using the Sheth-Tormen mass function \citep{Sheth:1999mn}.

The separation \ie\, $(s'_{\perp}, s'_{||})$ in the true cosmology might be different from those $(s_{\perp}, s_{||})$ in the fiducial cosmology, which is used to convert the redshifts to distances. This is known as the AP effect \citep{Alcock1979}, which can be accounted for via the following relation:
\begin{equation}
    s'_{\perp} = \alpha_{\perp}s_{\perp}, \quad\, s'_{||}=\alpha_{||}s_{||}\,.
\end{equation}
Here, two scaling factors $(\alpha_{\perp}, \alpha_{\parallel})$ are introduced to parameterise the differences of distances (across and along the LOS) between the true and fiducial cosmology:
\begin{equation}
\alpha_{\perp} =\frac{D_M(z)r_d^{\rm fid}}{D^{\rm fid}_M(z)r_d},\quad\,\alpha_{\parallel}= \frac{D_H(z)r_d^{\rm fid}}{D^{\rm fid}_H(z)r_d}\,,
\end{equation}
where $D_M(z)\equiv (1+z) D_A(z)$, and $D_A(z)$ is the angular diameter distance. $D_H(z)=c/H(z)$, $H(z)$ is the Hubble expansion parameter. The superscript `$\rm fid$' denotes the corresponding values in the fiducial cosmology. 

\begin{figure*}   
\includegraphics[scale=0.3]{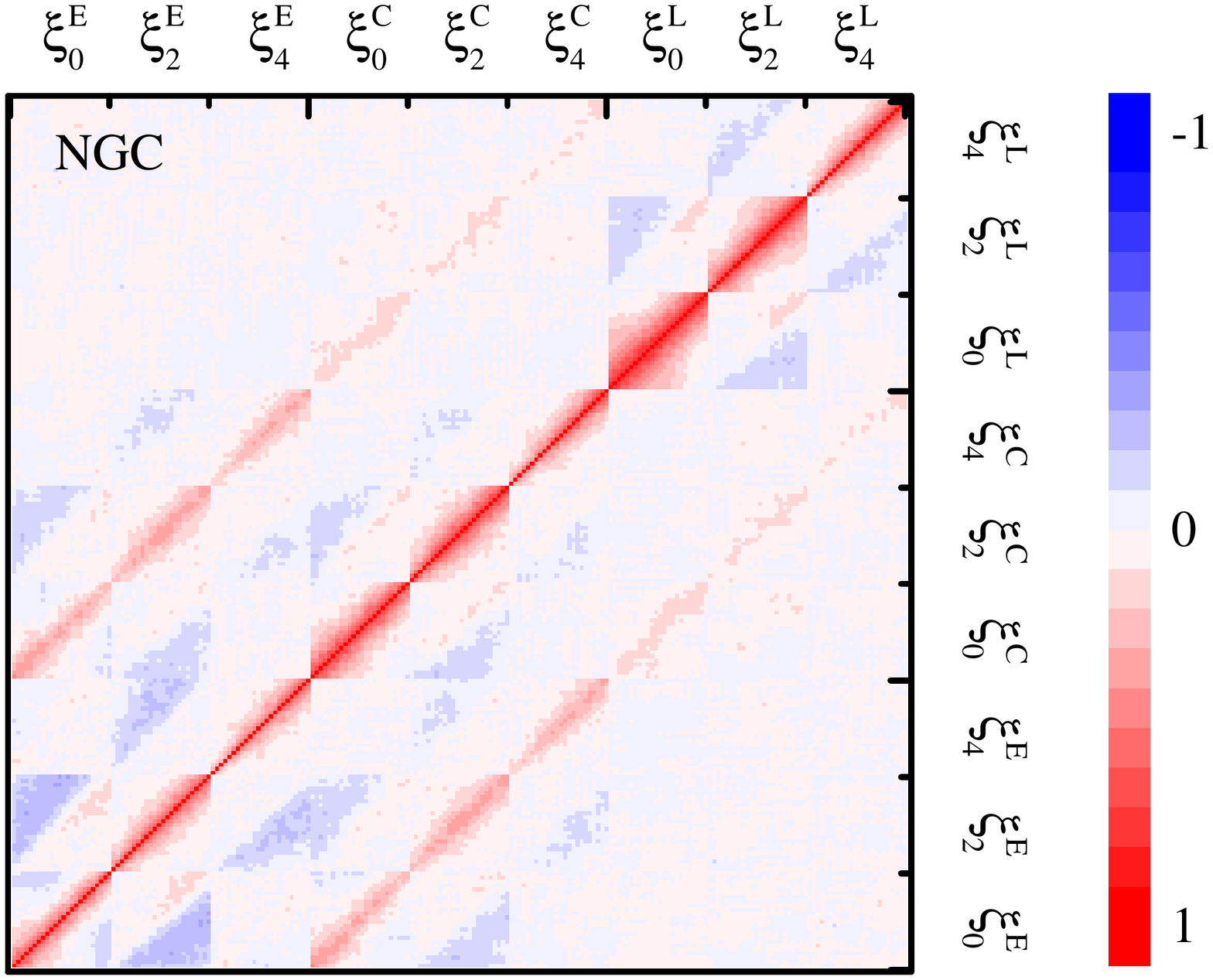}
\includegraphics[scale=0.3]{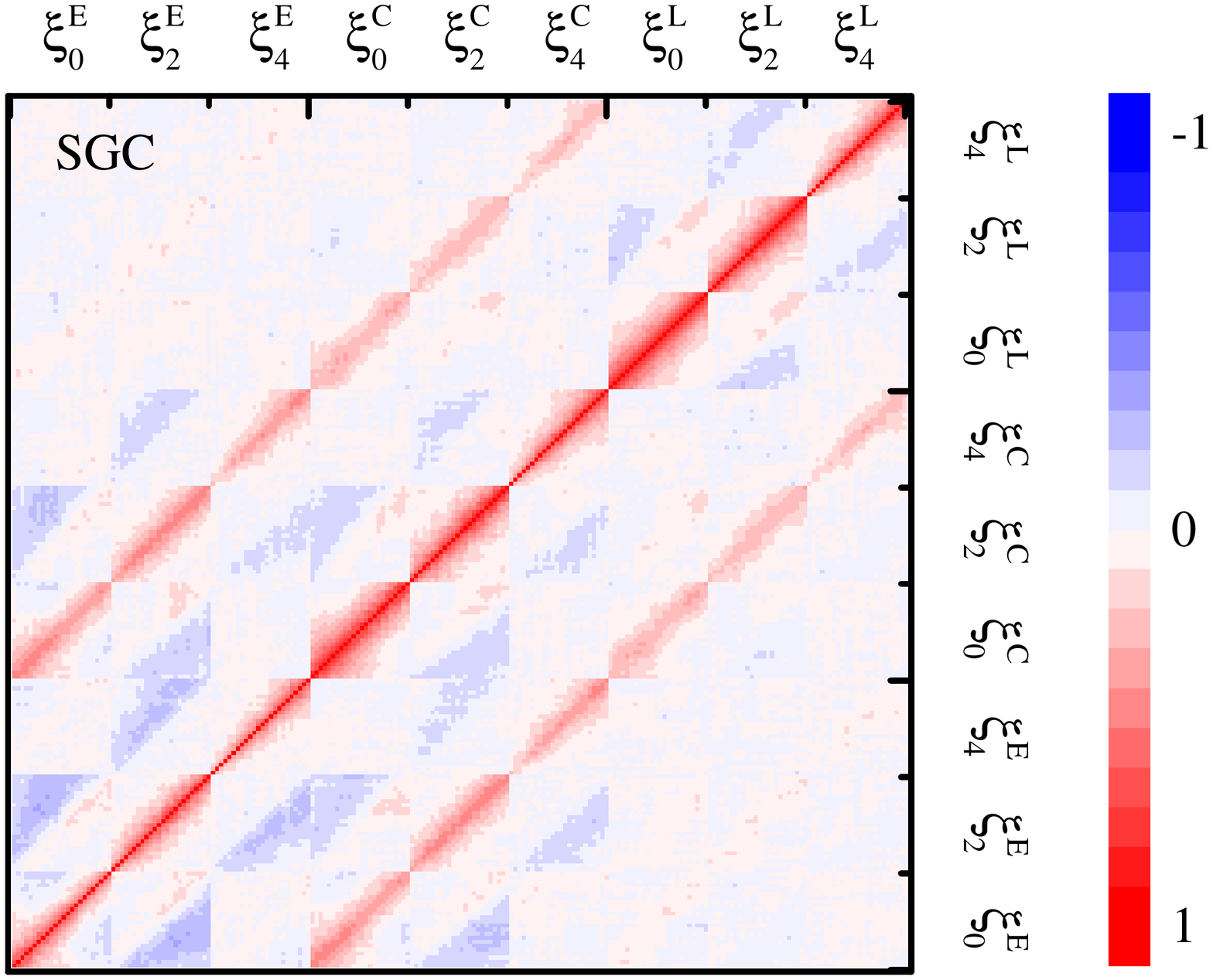}
\caption{The correlation matrices between the correlation function monopole, quadrupole, and hexadecapole measured from $1000$ EZmocks in the NGC (left) and SGC (right). For each measurement, $\xi^{\rm E}_\ell$, $\xi^{\rm C}_\ell$, or $\xi^{\rm L}_\ell$, we show the correlations for $24$ bins linearly even spaced in separation $s$ between $30$ to $150\, h^{-1} \rm Mpc$.}\label{fig:xilcorr} 
\end{figure*} 

\section{Measurements of correlation functions} 
\label{sec:measureCF}
We measure the auto-correlation functions for the ELG and LRGpCMASS samples using the \citet{LSestimator} estimator,
\begin{equation}
\label{eq:autoxi2D}
\xi(s,\mu) = \frac{DD(s,\mu)-2DR(s,\mu)+RR(s,\mu)}{RR(s,\mu)}\,,
\end{equation}
where $DD$, $DR$ and $RR$ are the weighted data-data, data-random and random-random pair counts with the separation $s$, and the cosine of the angle between the pair and the LOS, denoted as $\mu$. 

Additionally, we measure the cross-correlation between these two samples using the following estimator,
\begin{equation}
\label{eq:crossxi2D}
\xi(s,\mu) = \frac{D^{\rm E}D^{\rm L} -D^{\rm E}R^{\rm L} -D^{\rm L}R^{\rm E} +R^{\rm E}R^{\rm L} }{R^{\rm E}R^{\rm L} }\,,
\end{equation}
where superscripts `E' and `L' represent the ELG and LRGpCMASS samples, respectively.

The Legendre projections of the correlation function is calculated to obtain the correlation function multipoles,
\begin{equation}
    \xi_{\ell}(s) = \frac{2\ell+1}{2}\int_{-1}^{1}{\rm d}\mu\, \xi(s,\mu) \mathcal{L}_{\ell}(\mu)\,,
\label{eq:multipole_CF}
\end{equation}
where $\mathcal{L}_{\ell}(\mu)$ is the Legendre polynomial.

\begin{figure*}   
\includegraphics[scale=0.2]{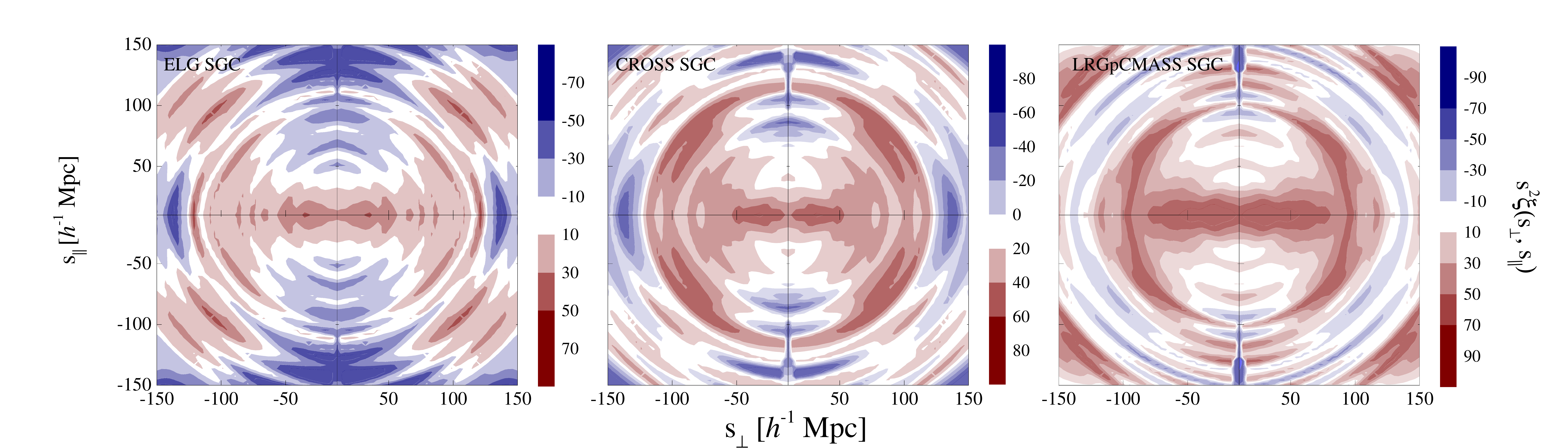}
\caption{The 2D correlation functions assembled using the measured monopole, quadrupole and hexadecapole, $i.e.$ $\xi(s,\mu) = \xi_0(s) \mathcal{L}_0(\mu)+ \xi_2(s) \mathcal{L}_2(\mu)+ \xi_4(s) \mathcal{L}_4(\mu)$, with $s^2= s_{\|}^2+ s_{\perp}^2$, from ELG SGC samples (left), LRGpCMASS SGC samples (right), and their cross-correlation (middle).
} 
\label{fig:xi2D}
\end{figure*} 

In Figure \ref{fig:xilMockCh}, we present measurements of the correlation function monopole and quadrupole, including the auto-correlation functions of LRG in blue (left panels) and ELG in green (right panels), and their cross-correlation in red (middle panels), using a set of MDPL2 mock with the $z$ LOS, which is produced via the multi-tracer HMQ HOD model. The correlation function multipoles are measured with a bin width of $5\,h^{-1} \rm Mpc$ within the scale range of $25-150\,h^{-1} \rm Mpc$. The error bar is estimated from the RascalC covariance matrix.

We show the correlation function multipoles measured from the DR16 galaxy samples and EZmocks in Figures \ref{fig:xilNGC} and \ref{fig:xilSGC} for measurements in the NGC and SGC, respectively. All the correlation function multipoles are measured with a bin width of $5\,h^{-1} \rm Mpc$ within the scale range of $30-150\,h^{-1} \rm Mpc$. The measurements of ELG are shown in upper panels, where the dashed-line and shaded areas display the $1\,\sigma$ regions evaluated from $1000$ ELG EZmocks without and with removing the RIC effect, respectively; The black-line areas are the mean of ELG EZmocks with the $1\,\sigma$ standard deviation after further removing the angular systematics using Eq.\,\ref{eq:modifiedGSM}; The black circles with the $1\,\sigma$ error bars are the multipoles measured from ELG samples with removing both the RIC effect and angular systematics in ELG data. 

In the middle panels of Figures \ref{fig:xilNGC} and \ref{fig:xilSGC}, we show measurements of cross-correlations between ELG and LRGpCMASS. The $1\,\sigma$ areas covered within the green dashed lines (RIC is not subtracted) and shaded regions (with RIC subtracted off) are evaluated from EZmocks. The green squares with the $1\,\sigma$ error bars are the measured multipoles from cross sample with the RIC effect removed. Within the $1\,\sigma$ region, the cross-correlation multipoles from EZmocks and data are mostly consistent on large scales. 

The panels in the bottom of Figures \ref{fig:xilNGC} and \ref{fig:xilSGC} are the measured multipoles from  LRGpCMASS sample and mocks. There is not much difference between blue dashed-line region (with RIC effect) and the blue shaded area (with removing RIC effect), which means that the RIC effect in LRGpCMASS data is negligible.  

The covariance matrix can be estimated using the measurements of $1000$ EZmocks,
\ba \label{eq:cov}
C^{\ell,\ell'}_{ij}=\frac{1}{N-1} \sum_{k=1}^{N} \left[\xi_{\ell}^k(s_i)-\bar{\xi}_{\ell}(s_i)\right] \left[\xi_{\ell'}^k(s_j)-\bar{\xi}_{\ell'}(s_j)\right],
\ea
where the average multipole is given by
\ba
\bar{\xi}_{\ell}(s_i)=\frac{1}{N} \sum_{k=1}^{N}\xi_{\ell}^k(s_i),
\ea
here $N=1000$ is the number of mock realisations. The normalised covariance matrices, \ie\,$C^{\ell,\ell'}_{ij}/\sqrt{C^{\ell,\ell}_{ii}\times C^{\ell',\ell'}_{jj}}$, in NGC and SGC are displayed in the left and right panels of Figure \ref{fig:xilcorr}, respectively. We fit $\xi_{\ell} (\ell=0,2,4)$ in the range $30<s<150 \,h^{-1} \rm Mpc$ ($72$ data points for each sample). The matrix contains $72\times72$ $s$ bins for each tracer, and so totally there is a $216\times216$ covariance matrix for the combined data vector of two auto-correlation and one cross-correlation measurements.

We show the 2D correlation function reconstructed from the measured monopole, quadrupole and hexadecapole using the ELG, LRGpCMASS, and their cross samples in SGC in Figure \ref{fig:xi2D}, where the BAO ring at $\sim 100\,h^{-1} \rm Mpc$ and the squashing effect due to RSD is clearly observed. 

We quantify the Signal-to-Noise Ratio (SNR) of measurement on the cross-correlation between two tracers of eBOSS via
\ba
\left(\rm SNR\right)^2_{\xi_{\ell}^{\rm C}} = \sum_{i,j} \left[ \xi_{\ell}^{\rm C} (s_i)\right]^T F^{\rm C}_{ij} \left[ \xi_{\ell}^{\rm C} (s_j)\right] \,,
\ea
where $F^{\rm C}_{ij}$ is the inverse covariance matrix for the measured cross-correlation. We obtain a detection of the cross-correlation function at a significance of $15\, \sigma$.

\section{Parameter estimation} 
\label{sec:BAORSDfit}
We perform a global fitting in the following parameter space, \ie\,
\begin{eqnarray}
{\bf p} \equiv \left\{\alpha_{\perp}, \alpha_{||}, b^m_{\rm NGC} \sigma_8, b^m_{\rm SGC} \sigma_8, f \sigma_8, \sigma^m_{\rm FoG}\right\} \,,
\end{eqnarray}
where $m={\rm E}, {\rm L},$ or C when using the ELG, LRGpCMASS, or CROSS sample alone. We use different bias parameters for NGC and SGC. Namely for the fit to each sample, we have $N_p=6$ free parameters in each case. 

For the combined fits of two samples, \eg\, ELG + LRGpCMASS, the free parameters for bias factors are
\begin{eqnarray}
\left\{b^m_{\rm NGC} \sigma_8, b^m_{\rm SGC} \sigma_8 \right\} &=& \left\{b^{\rm E}_{\rm NGC} \sigma_8, b^{\rm L}_{\rm NGC} \sigma_8, b^{\rm E}_{\rm SGC} \sigma_8,b^{\rm L}_{\rm SGC} \sigma_8 \right\} \nonumber \\ 
\left\{\sigma^m_{\rm FoG}\right\} &=& \left \{\sigma^{\rm E}_{\rm FoG},\sigma^{\rm L}_{\rm FoG} \right\} \,.
\end{eqnarray}
In total, we have $N_p=9$ free parameters. The number of free parameters in the cases of ELG+CROSS and CROSS+LRGpCMASS are also $N_p=9$. 
For the joint fit of ELG, LRGpCMASS and cross samples, as the bias of the cross sample can be derived from the biases of ELG and LRGpCMASS via Eq. \ref{eq:biasCross}, in principle we do not need to introduce additional degrees of freedom for the bias factors for the cross sample. We only assign a new damping parameter, \ie\, $\sigma^{\rm C}_{\rm FoG}$ to the cross-correlation function, thus we have $N_p=10$ free parameters for the joint fit. However, considering that the LRGpCMASS and ELG samples of eBOSS DR16 are not fully overlapping, we also implement a fit by additionally introducing a set of bias parameters, \ie\,$\left\{b^{\rm C}_{\rm NGC} \sigma_8, b^{\rm C}_{\rm SGC} \sigma_8 \right\}$ for the cross sample. In this case the number of free parameters is $N_p=12$. 

We use a modified version of {\tt CosmoMC}\footnote{\url{http://cosmologist.info/cosmomc/}} \citep{cosmomc} based on a Markov Chain Monte Carlo (MCMC) technique to sample the parameter space $\bold{p}$, and search for the minimum $\chi^2$,
\ba
 \chi^2 (\bold{p}) = \chi_{\rm NGC}^2 (\bold{p}|b_{\rm NGC} \sigma_8) + \chi_{\rm SGC}^2 (\bold{p}|b_{\rm SGC} \sigma_8) \,,
\ea
where
\ba
\chi_{\rm NGC}^2 (\bold{p}|b_{\rm NGC} \sigma_8) \equiv 
 \sum_{i,j}^{\ell,\ell'}  \left[\xi^{th}_{\ell} (s_i, \bold{p}|b_{\rm NGC} \sigma_8) -\xi_{\ell}(s_i) \right] \nonumber \\
 F^{\ell,\ell'}_{ij,{\rm NGC}} \left[\xi_{\ell'}^{th}(s_j, \bold{p}|b_{\rm NGC} \sigma_8) -\xi_{\ell'}(s_j)\right]\,,
\ea
for observed multipoles $\{\xi_\ell(s_i)\}$ and 
\ba
\chi_{\rm SGC}^2 (\bold{p}|b_{\rm SGC} \sigma_8) \equiv 
 \sum_{i,j}^{\ell,\ell'}  \left[\xi^{th}_{\ell} (s_i, \bold{p}|b_{\rm SGC} \sigma_8) -\xi_{\ell}(s_i) \right] \nonumber \\
 F^{\ell,\ell'}_{ij,{\rm SGC}} \left[\xi_{\ell'}^{th}(s_j, \bold{p}|b_{\rm SGC} \sigma_8) -\xi_{\ell'}(s_j)\right]\,,
\ea
here $F^{\ell,\ell'}_{ij}$ is the inverse of the covariance matrix in Eq.\ref{eq:cov}. An unbiased estimation for the inverse covariance matrix is given by
\ba \label{eq:inv_cov}
\widetilde{C}_{ij}^{-1}=\frac{N-N_b-2}{N-1}C_{ij}^{-1}.
\ea
where $N_b$ is the number of bins. In order to include the error propagation from the error in the covariance matrix into the fitting parameters \citep{Percival2014} we rescale the covariance matrix, $\widetilde{C}_{ij}$, by 
\ba
M=\frac{1+B(N_b-N_p)}{1+A+B(N_p+1)}\,,
\ea
here $N_p$ is the number of the fitting parameters, and 
\ba
A=\frac{2}{(N-N_b-1)(N-N_b-4)}\,, \\
B=\frac{N-N_b-2}{(N-N_b-1)(N-N_b-4)} \,.
\ea

\begin{figure*}   
\includegraphics[scale=0.6]{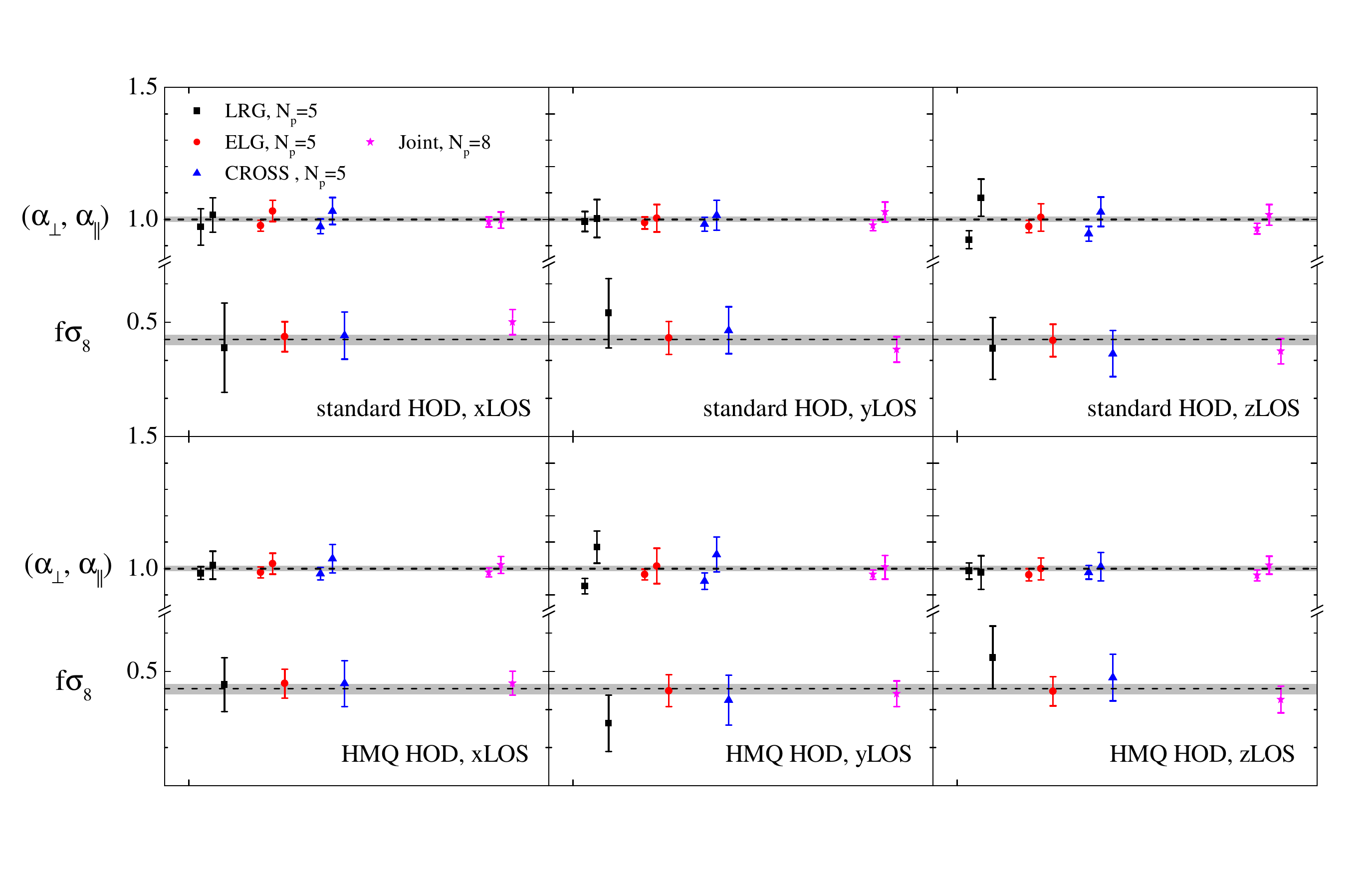}
\caption{The best-fits from measurements of LRG and ELG mock galaxy catalogues, and their cross correlation using MDPL2 mocks in \citet{Alam:2019pwr}. The shaded bands show an error of $1\%$ on the $\alpha_{\perp}$ and $\alpha_{||}$ parameters and $3\%$ on $f\sigma_8$, and the dashed lines in the middle of the shaded area are the fiducial parameter values. The multi-tracer MDPL2 mock has two types of HOD models, $i.e.$ standard (upper panels) and HMQ (lower panels) with the LOS of $x,y,z$, so we have six realisations in total. We fit the LRG auto-correlation (black), ELG auto-correlation (red), and their cross-correlation (blue), respectively, and then perform a joint fitting using these three sets of measurements (magenta). Note that for the MDPL2 mock sample, we do not need to assign bias parameters for NGC and SGC separately, thus the number of free parameters for the fit of each sample is $N_p=5$, with $N_p=8$ for the joint fit.
}\label{fig:MockChallfit}
\end{figure*} 

\begin{figure*}   
\includegraphics[scale=0.3]{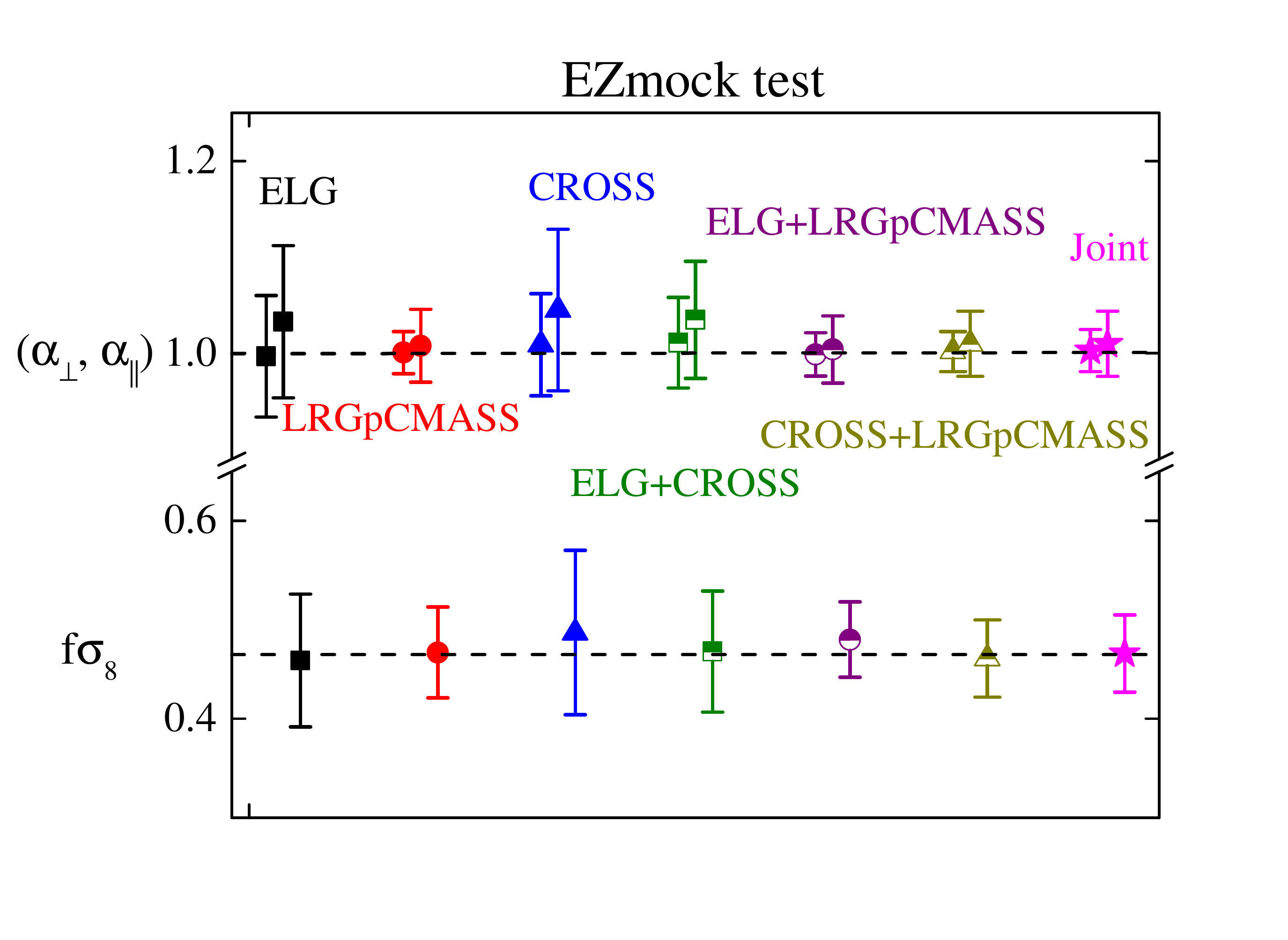}
\includegraphics[scale=0.3]{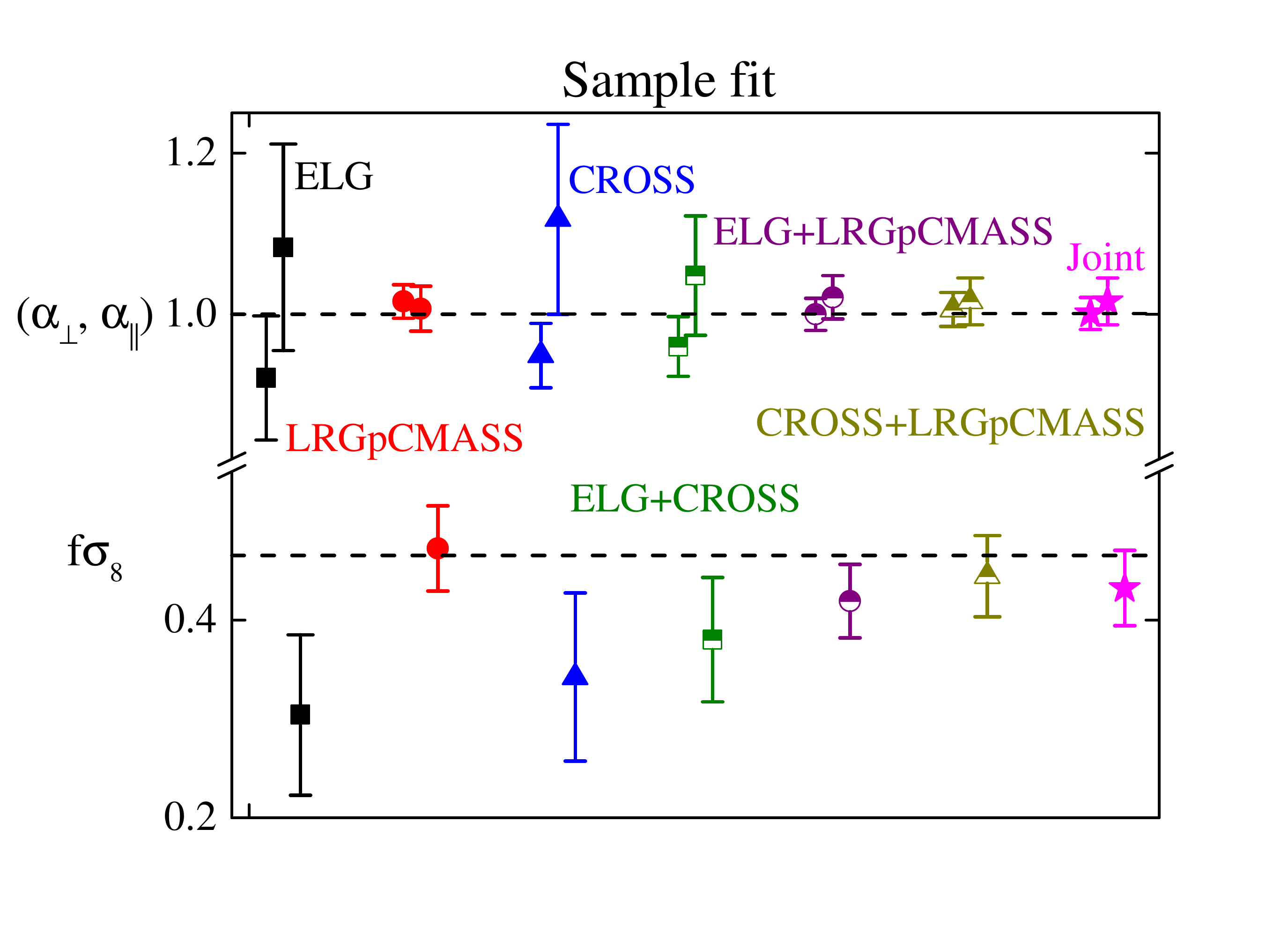}
\caption{The mean values with $1\,\sigma$ error bars from EZmock tests (left) and data fits (right) of different combinations, as shown in the legend.}\label{fig:fit_error}
\end{figure*} 

\section{Mock tests} 
\label{sec:mockresult}

We validate our pipeline in this section, using two series of mock catalogues, namely, the N-body MDPL2 mocks and $1000$ realisations of the EZmocks, as introduced in Sec. \ref{sec:mock}.

\subsection{MDPL2 mock fits} 

Figure \ref{fig:MockChallfit} shows the $\alpha_{\perp}, \alpha_{||}$ and $f\sigma_8$ parameters fitted to the MDPL2 mock. The multi-tracer MDPL2 mock has two types of HOD models, \ie\, standard (upper panels) and HMQ (lower panels), and we consider that the LOS is along $x,y$ or $z$ axis, so we have six realisations in total. We perform the fit to LRG auto-correlation, ELG auto-correlation, and their cross-correlation. The corresponding constraints on the $\alpha_{\perp}, \alpha_{||}$ and $f\sigma_8$ parameters from these three sets of measurement are displayed in black, red, and blue, respectively. The fitted results are generally within the error of $1\%$ for $\alpha_{\perp}$ and $\alpha_{||}$, and the error of $3\%$ for $f\sigma_8$. Following this, we perform a joint fit to these three sets of measurements together. The fitted results (magenta in Figure \ref{fig:MockChallfit}) are consistent with the expected values of the $\alpha_{\perp}, \alpha_{||}$ and $f\sigma_8$ parameters. 

\begin{table*}
\caption{The result of the fit to the mean of $1000$ EZmocks. $\Delta(p)$ shows the difference between the mean value from mock test and its expected value. The expected values of  $\alpha_{\perp}$ and $\alpha_{\parallel}$ are $1$. The expected values of $f\sigma_8$ at different $z_{\rm eff}$ are $f\sigma_8(z_{\rm eff}=0.70)=0.471\,,f\sigma_8(z_{\rm eff}=0.77)=0.465\,,f\sigma_8(z_{\rm eff}=0.845)=0.458$. }
\begin{center} 
\begin{tabular}{ccccc}
\hline\hline
Catalogues & $z_{\rm eff}$  &$\Delta(\alpha_{\perp})$ &$\Delta(\alpha_{\parallel})$ &$\Delta(f \sigma_8)$ \\ \hline
ELG & $0.845$ & $-0.001 \pm 0.061 $ &  $0.029 \pm 0.076$ &  $\,\,\,\,0.003 \pm 0.066$ \\
ELG  & $0.770$ & $-0.003 \pm 0.063 $ &  $0.033 \pm 0.079 $ & $-0.006 \pm 0.067$ \\ \hline
LRGpCMASS &	$0.700$	& 	$ 0.001 \pm 0.022$ & $0.010 \pm 0.037$	&	$-0.002	\pm	0.045$ \\
LRGpCMASS &	$0.770$	& 	$ 0.001 \pm 0.022$ & $0.008 \pm 0.038$	&	$\,\,\,\,0.002	\pm	0.046$ \\ \hline 
CROSS & $0.770$ & $0.009 \pm	0.053$	&	$0.045 \pm 0.084$	&	$\,\,\,\,0.022	\pm	0.083$ \\ 
\hline 
ELG+CROSS & $0.770$ & $0.011 \pm 0.047 $ &  $0.035 \pm 0.061 $ & $\,\,\,\, 0.003 \pm 0.061$ \\
ELG+LRGpCMASS & $0.770$ & $0.002 \pm 0.021 $ & $0.010 \pm 0.034 $ & $-0.004 \pm 0.039 $ \\ 
CROSS+LRGpCMASS & $0.770$ & $0.003 \pm 0.022 $ & $0.009 \pm 0.036 $ & $\,\,\,\,0.010 \pm 0.044 $ \\ \hline
Joint ($N_p=10$) & $0.770$ & $ 0.002 \pm  0.022$ &  $ 0.007 \pm 0.034 $ & $ -0.001\pm 0.037 $ \\ 
Joint ($N_p=12$) & $0.770$ & $0.003 \pm 0.022$ &  $0.010 \pm 0.034$& $\,\,\,\,0.001 \pm 0.039 $ \\ 
\hline\hline                      
\end{tabular}
\end{center}
\label{tab:mock_test}
\end{table*}

\begin{figure}   
\includegraphics[scale=0.4]{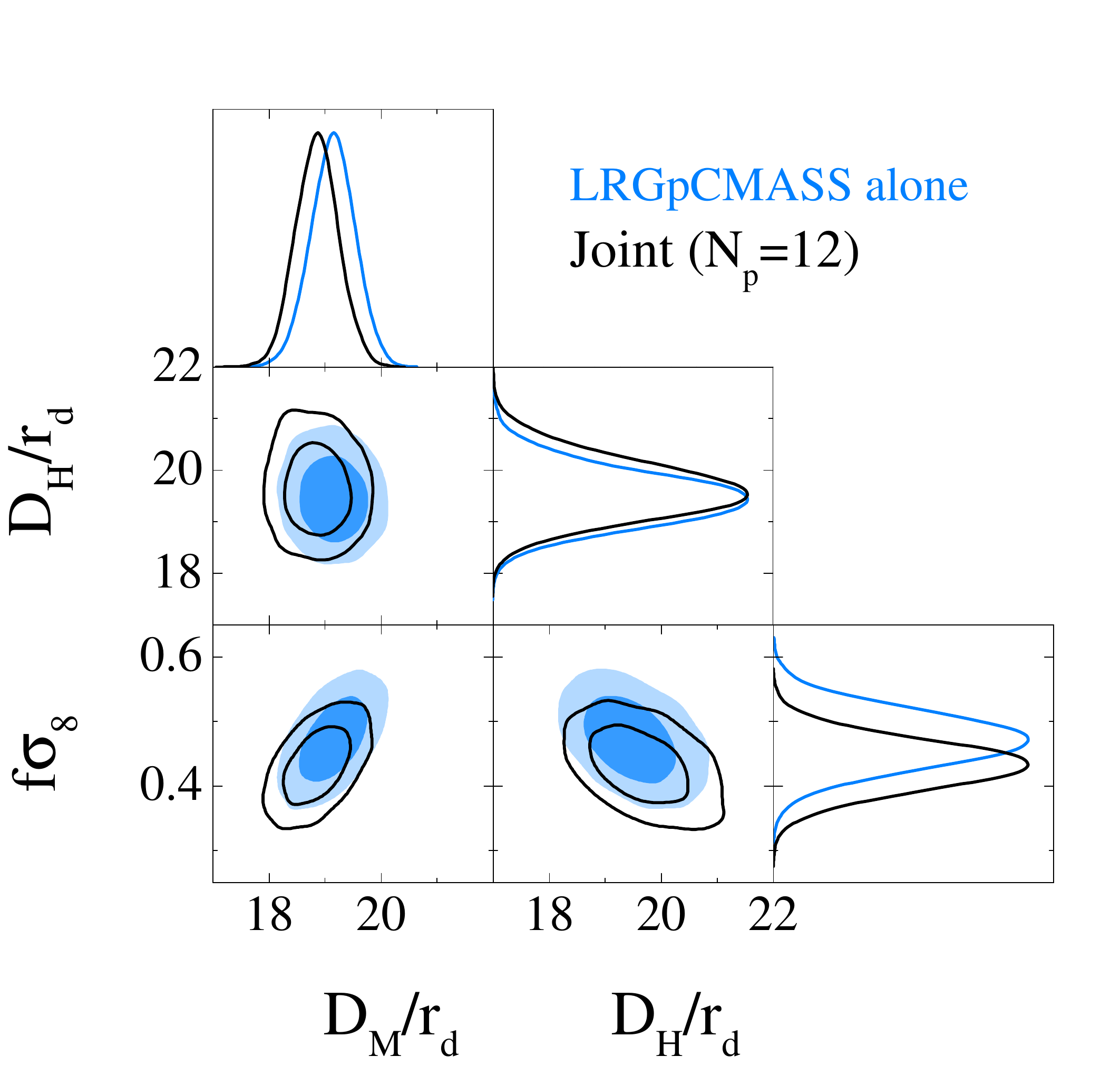}
\caption{The one-dimensional posterior distributions and the 68 and 95 \% CL contour plots for the $D_M/r_d$, $D_H/r_d$, and $f\sigma_8$ parameters using LRG samples alone (blue), and the joint constraint (black). }\label{fig:contour_bao_rsd}
\end{figure} 

\subsection{EZmock tests} 
We apply our pipeline to the average of the correlation function multipoles, measured from $1000$ realisations of the EZmocks, and present the marginalised mean values with $68\%$ CL uncertainty of BAO and RSD parameters in Table \ref{tab:mock_test} and in the left panel of Figure \ref{fig:fit_error}. As detailed previously, the ELG, LRGpCMASS and their cross correlation can be best modelled at effective redshifts of $0.845, 0.7$ and $0.77$, respectively, but for the joint fit, we make an assumption that all three correlation functions can be modelled using a fixed template at $z_{\rm eff}=0.77$, which is explicitly tested here.

\begin{figure*} 
\centering
\includegraphics[scale=0.2]{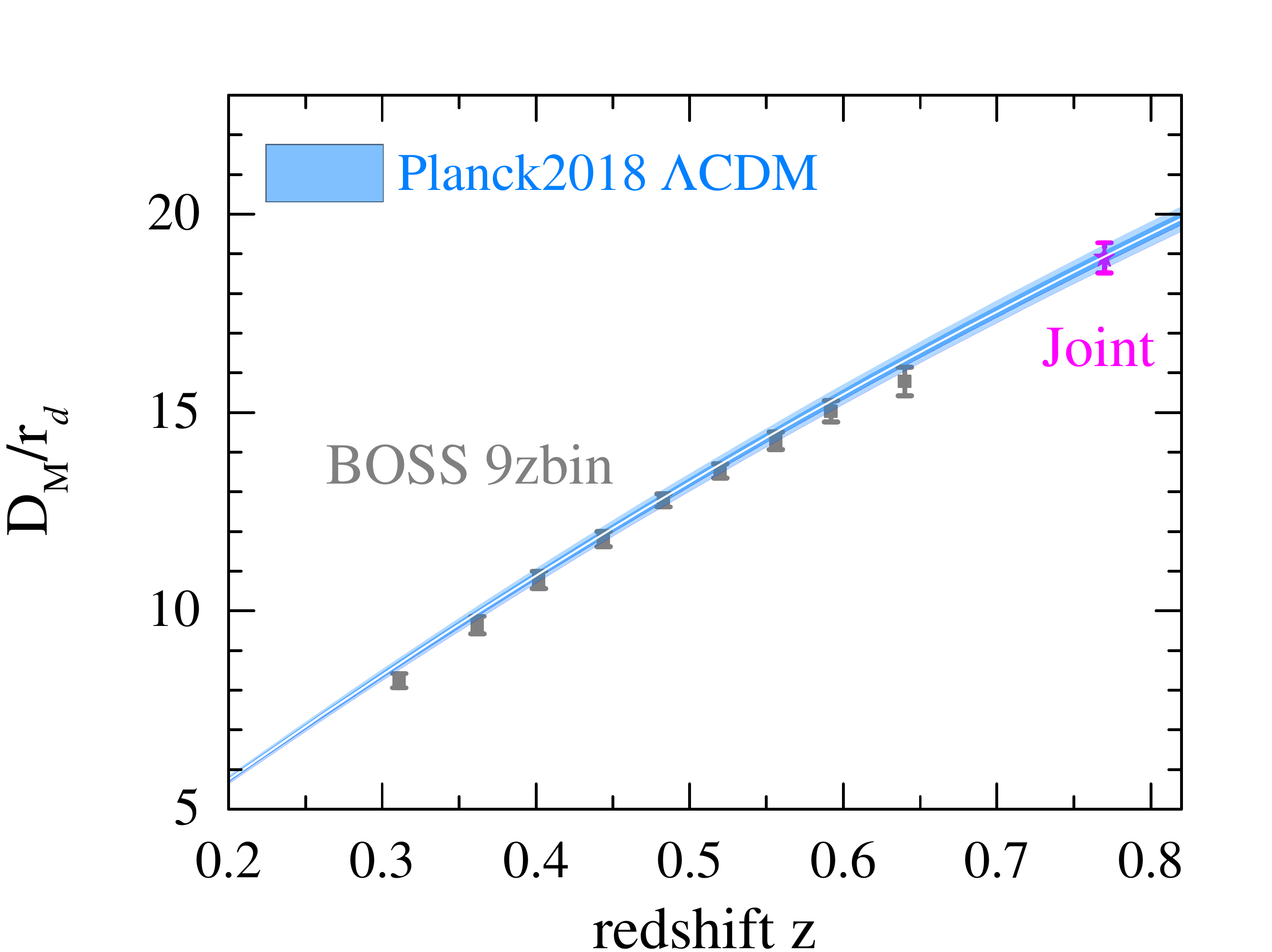} 
\includegraphics[scale=0.2]{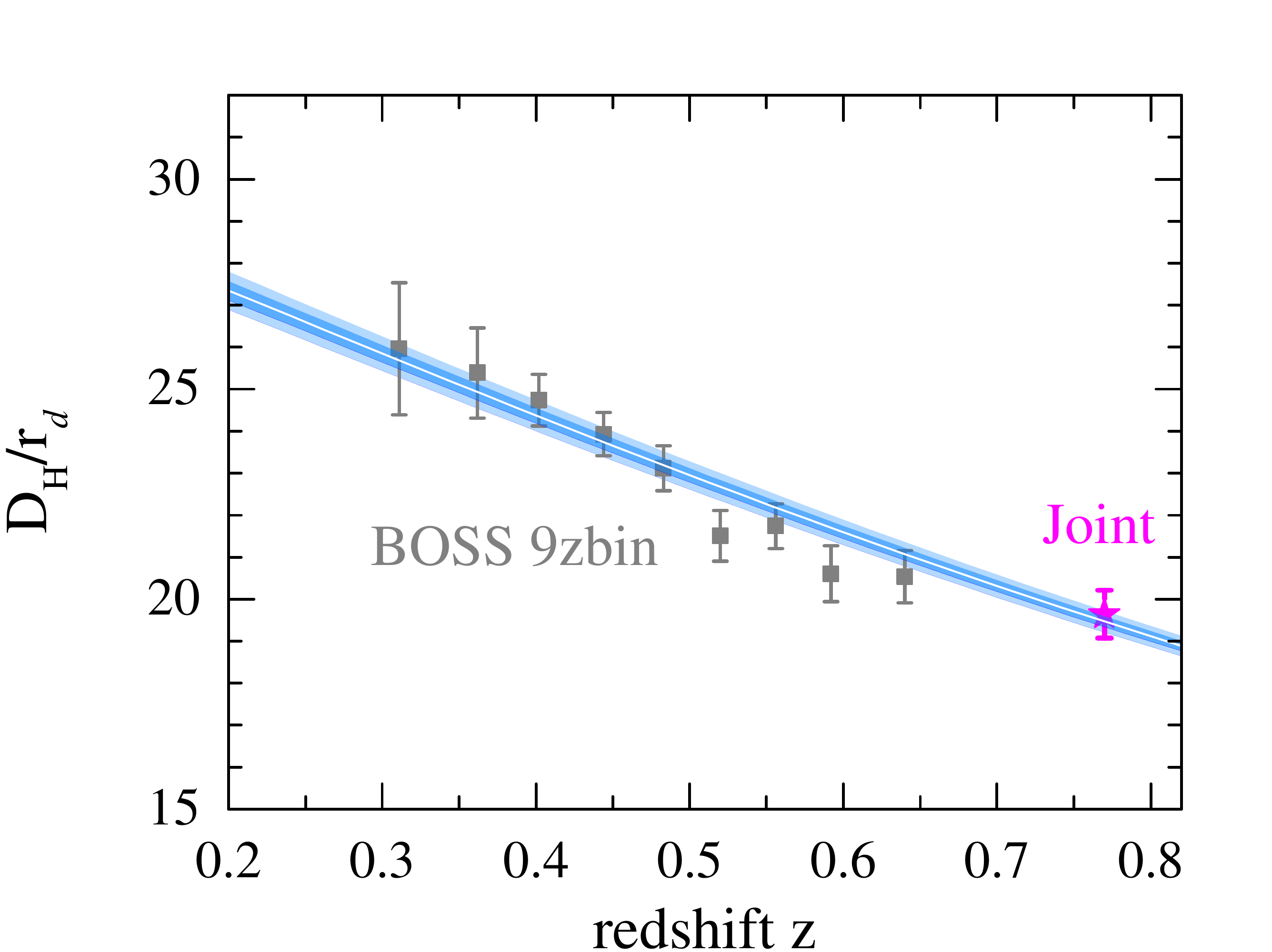} 
\includegraphics[scale=0.2]{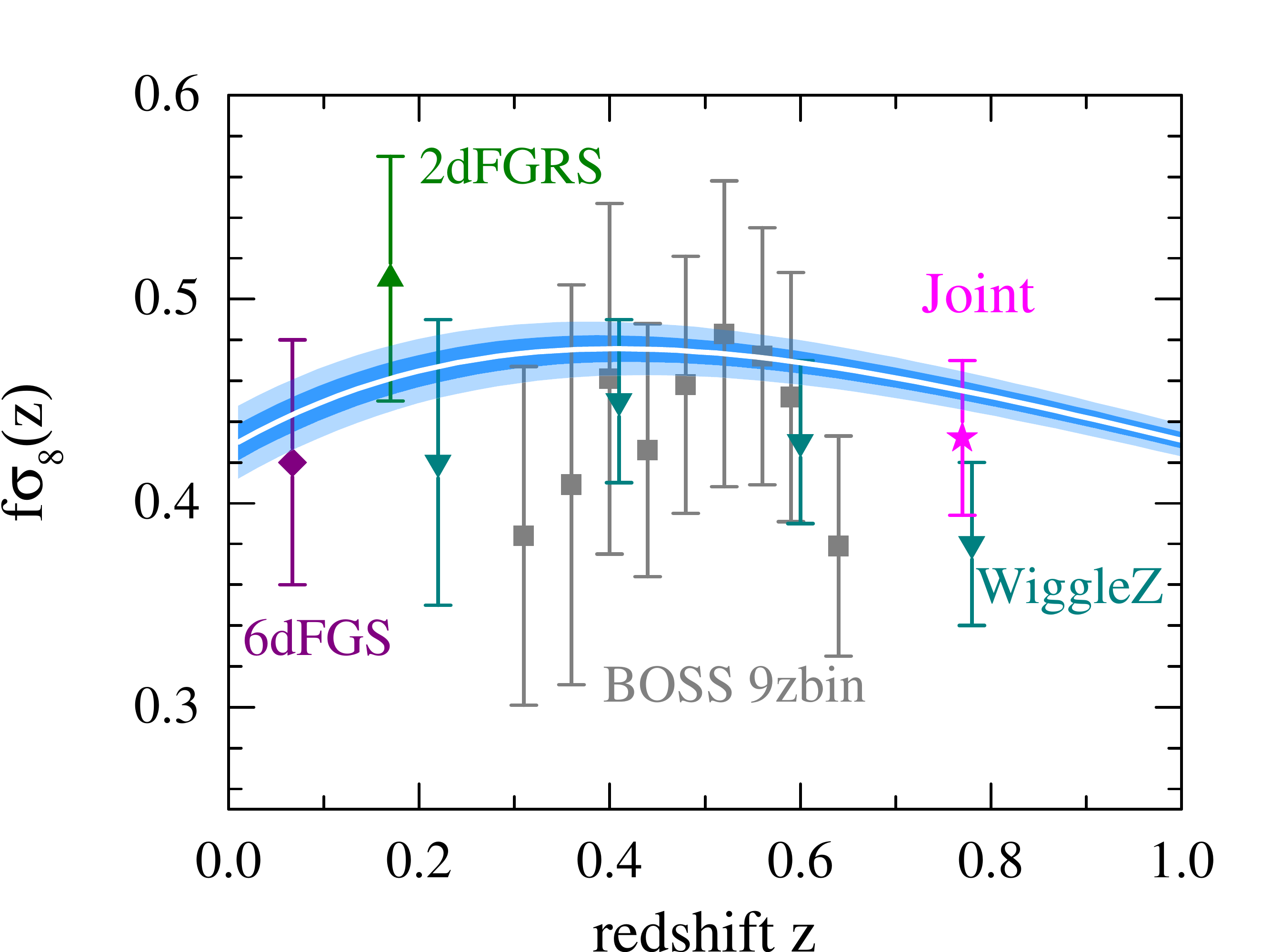} 
\caption{The evolution of $D_M/r_d$, $D_H/r_d$, and $f\sigma_8$ as a function of $z$. For reference, the blue bands are the predictions from Planck 2018 in the $\Lambda$CDM cosmology \citep{PLC2018}.}
\label{fig:PLC_bao_rsd}
\end{figure*}

As shown in Table \ref{tab:mock_test}, the observables of each tracer can be well fitted by a template created at their corresponding effective redshifts, and the bias of the fitting is well within 68\% CL. We then proceed to tests of all observables at $z_{\rm eff} =0.77$, and find almost no change on the posterior of parameters. This demonstrates that it is reasonable to model all three sets of observables at $z_{\rm eff} =0.77$, which is the effective redshift of the cross correlation. The joint fitting at $z_{\rm eff} =0.77$ successfully returns the input values of parameters with a marginal bias, which further validates our pipeline. 

\section{Data fits} 
\label{sec:dataresult}
We present measurements of the BAO and RSD parameters from the DR16 samples in Table \ref{tab:data_fit} and in right panel of Figure \ref{fig:fit_error}, and find consistent BAO and RSD measurements from LRGpCMASS sample with the fiducial cosmology given their statistical uncertainties. Compared to results of the single-tracer analysis, the measurements of BAO and RSD from cross alone is consistent with ELG sample within the $1\,\sigma$ error bar. The LRGpCMASS gives a much smaller statistical uncertainty than that of ELG. The difference between $f\sigma_8$ values from LRGpCMASS and cross sample is $1.11\,\sigma$.

Combining the ELG, LRGpCMASS, and cross samples, \eg\, $\rm ELG+CROSS, ELG+LRGpCMASS$, or $\rm CROSS+LRGpCMASS$, we obtain improved constraints. These measurements are fully consistent within $1\,\sigma$ error.

The joint fits from ELG and LRGpCMASS auto-correlation functions and their cross correlation give the tightest constraints. For joint fits, we present the results in two cases, \ie\, $N_p=10$ denotes that we did not assign additional bias parameters for the cross samples, which are derived via Eq.\ref{eq:biasCross}; $N_p=12$ means that the cross sample has its own bias parameters. We find the BAO and RSD measurements in these two joint cases are in good agreement. Comparing with the fitted result from LRGpCMASS alone, we find the Figure of Merit (FoM) of the $\alpha_{\perp}, \alpha_{||}$ and  $f\sigma_8$ parameters, $\rm FoM=1/\sqrt{\rm det\,Cov(\alpha_{\perp}, \alpha_{||}, f\sigma_8)}$, from the joint ($N_p=12$) fit is improved by a factor of $1.11$. In particular, the improvement in the measurement precision of $f\sigma_8$ is $11.6$\% over that using only the LRGpCMASS sample.

We also perform an analysis when the AP parameters are fixed to $1$, as a consistency test of the fiducial cosmology. As expected, we get a tighter constraint on $f\sigma_8$ in this case, namely, the statistical uncertainty of $f\sigma_8$ with AP fixed is reduced by $\sim (24\% - 49\%)$ compared with results with AP parameters marginalized over. In cases with AP parameters fixed, we obtain a $9\%$ improvement in the statistical precision of $f\sigma_8$ from the joint fit compared with the LRGpCMASS's constraint. We compare our result on $f\sigma_8$ with AP parameters fixed with the forecast published in \citep{eBOSSZhao16}, where the AP parameters are also fixed. Because the actual survey area is different from that used in the forecast, and the error on parameters is inversely proportional to the square root of the survey area, we preform a rescaling of the forecast using the areas, and find that the improvement on the precision of $f\sigma_8$ is 14\%, which is slightly better than our actual analysis. 

\begin{figure*} 
\centering
\includegraphics[scale=0.28]{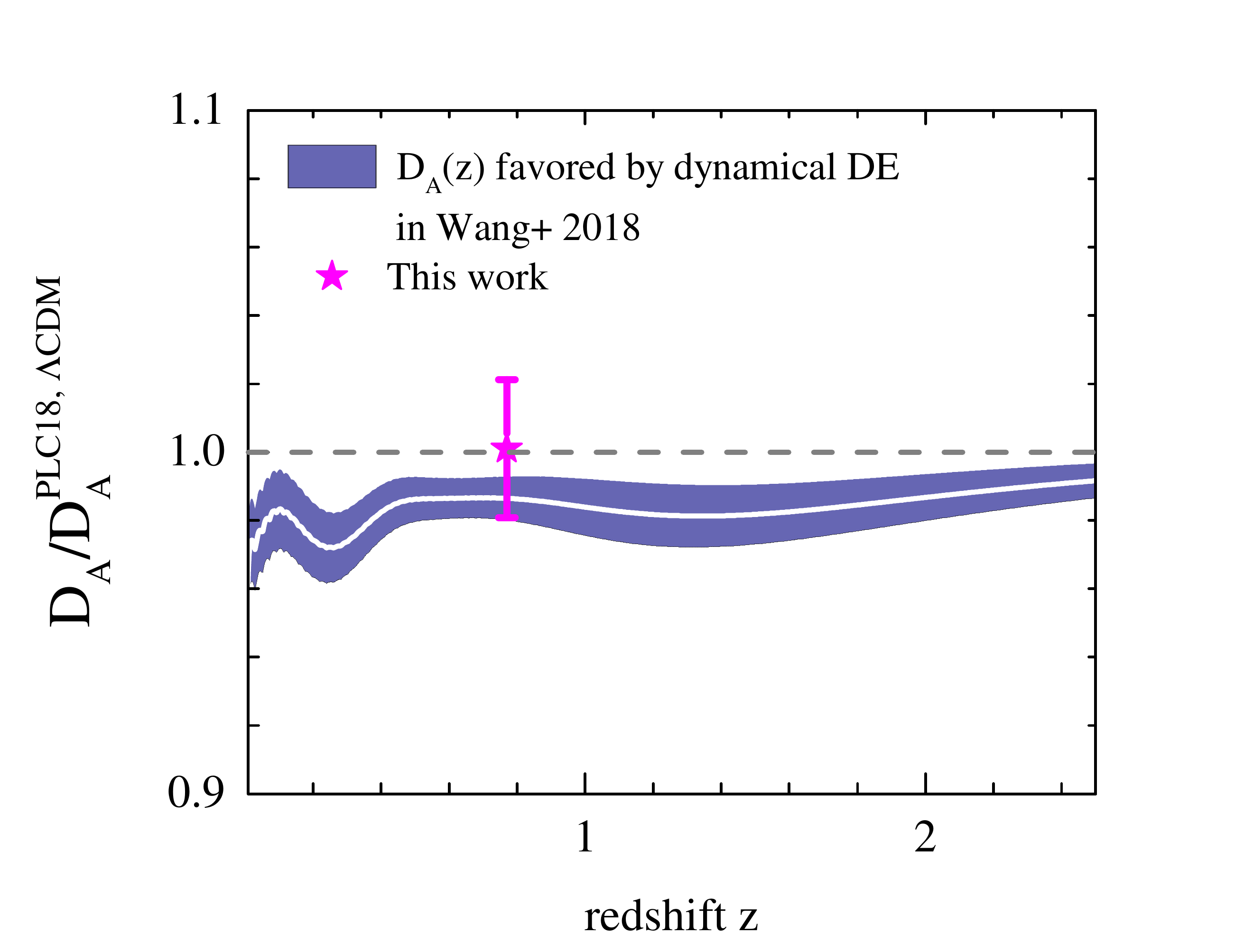} 
\includegraphics[scale=0.28]{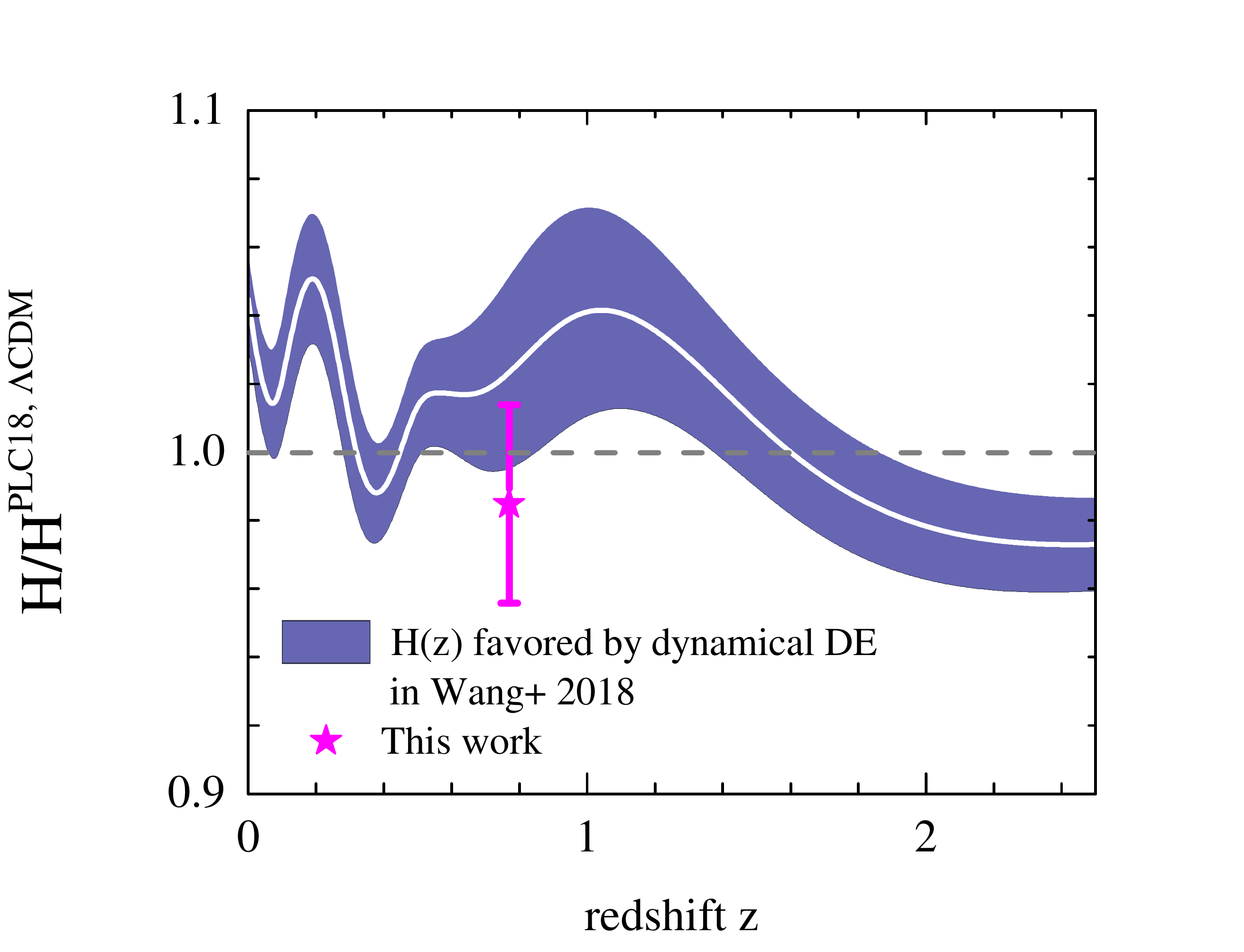} 
\caption{The shaded bands are the uncertainties of angular diameter distance, $D_A(z)$ (left) and Hubble expansion rate, $H(z)$ (right)  favored by the reconstructed dynamical dark energy in \citep{Wang:2018fng}. The data point with error bar is our measurement in this work. They are rescaled by the mean values in the $\Lambda$CDM predicted by Planck 2018 \citep{PLC2018}.}
\label{fig:compare_DDE}
\end{figure*}

\begin{figure}   
\includegraphics[scale=0.35]{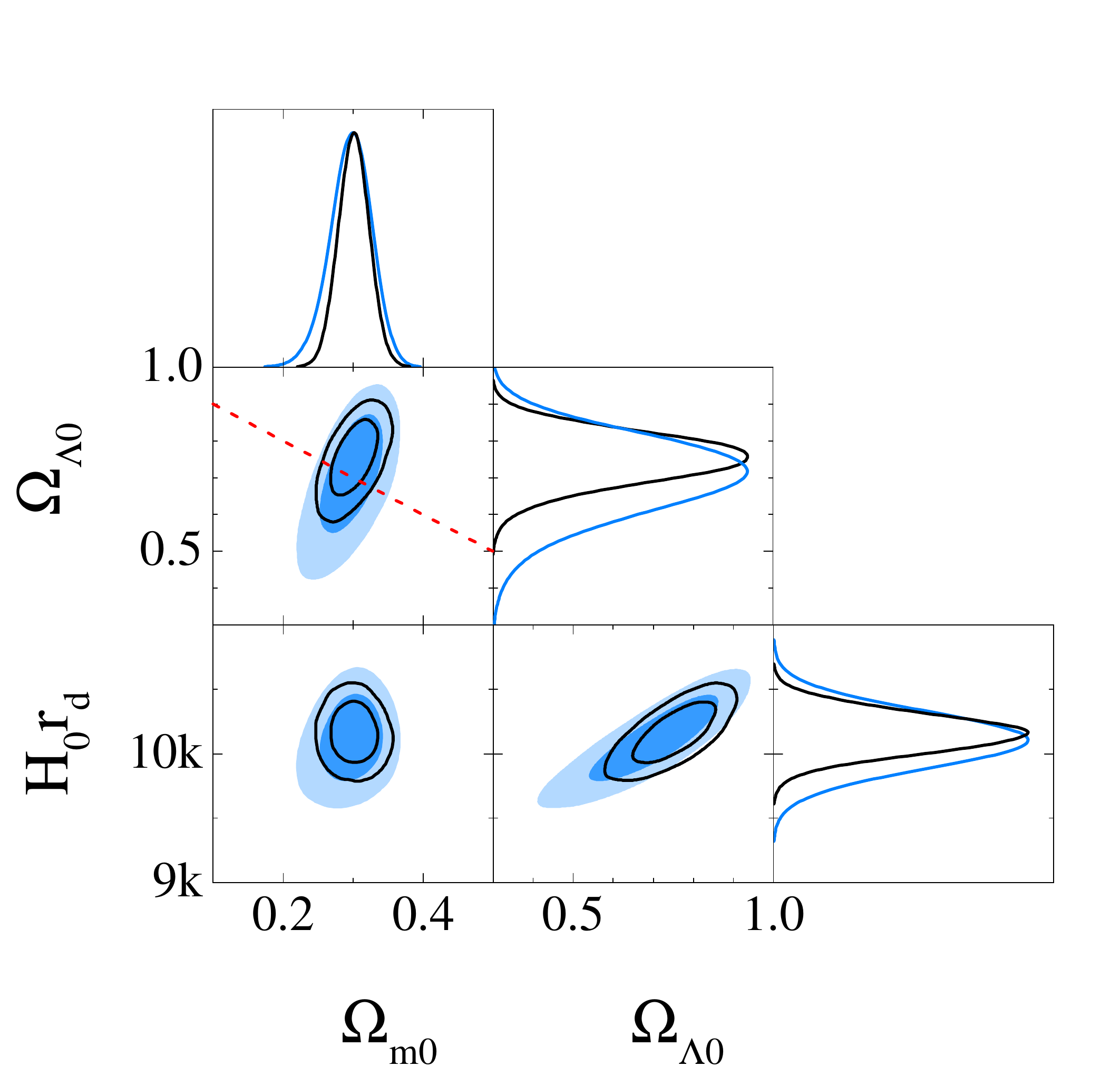}
\caption{The one-dimensional posterior distributions and the 68 and 95 \% CL contour plots for the cosmological parameters using MGS \citep{Ross:2014qpa} + 6dFGRS \citep{6dFBAO} +BOSS DR12 (low-$z$ and middle-$z$ bins) \citep{Alam2016} + eBOSS DR16 QSO \citep{Neveux2020, Hou2020} + eBOSS DR16 Lyman-$\alpha$ forests \citep{Bourboux2020} + our joint ($N_p=12$) result (black), compared with the constraining result (blue) in eBOSS DR14 paper \citep{Ata:2017dya}. The red dashed line represents a model with zero curvature.}\label{fig:contour_OMOL}
\end{figure} 

We derive the parameters $D_M/r_d = 18.86 \pm 0.38$ and $D_H/r_d = 19.64 \pm 0.57$ from the joint ($N_p=12$) fitted results on $\alpha_{\perp}$ and $\alpha_{||}$ in Table \ref{tab:data_fit}. The 1D posterior distributions of $D_M/r_d$, $D_H/r_d$, and $f\sigma_8$, and their 2D contour plots from the LRGpCMASS alone (blue) and the joint fit (black) are shown in Figure \ref{fig:contour_bao_rsd}. 

For the joint fits, the best fit values and covariance matrix for the ($D_M/r_d$, $D_H/r_d$, $f\sigma_8$) parameters are given by
  \begin{equation}
  \label{eq:vector}
{\bf D} \equiv
 \begin{pmatrix}
D_M/r_d \\
D_H/r_d \\
f\sigma_8 \\
 \end{pmatrix}= 
  \begin{pmatrix}
18.86  \\
19.64  \\
0.432
 \end{pmatrix},
 \end{equation}
and 
 \begin{equation} 
 \label{eq:paracov}
C = 10^{-3}
 \begin{pmatrix}
141.0707 & -15.7168 & 7.5252 \\
 & 321.7959 & -9.5995 \\
 & & 1.4812
 \end{pmatrix} 
\end{equation}
which are used in the cosmological implications section. 

We recommend users to use the joint measurement\footnote{The multi-tracer BAO and RSD measurements and covariance matrix are available at \url{https://github.com/ytcosmo/MultiTracerBAORSD/}. This measurement can be used together with the BAO and RSD measurements in the first six $z$ bins $i.e.$ $0.2<z<0.59$ from BOSS DR12 in \citep{Wang:2017wia}.} reported in Eqs. \ref{eq:vector} and  \ref{eq:paracov} to perform constraints on dark energy or tests of gravity. 

In Figure \ref{fig:PLC_bao_rsd}, we present our BAO and RSD measurements alongside the $\Lambda$CDM prediction from Planck 2018 \citep{PLC2018}. Our measurement is consistent with these predictions. 

We also show our BAO and RSD measurements and the BAO distances favored by the reconstructed dynamical dark energy from a combined observational data \citep{Wang:2018fng} together in Figure \ref{fig:compare_DDE}. There is no significant tension between the new measurement and the prediction of the reconstructed dynamical dark energy within $1\,\sigma$ statistical error, although the measurement is more consistent with Planck 2018. 

\begin{table*}
\centering
\caption{The mean values with 68\% CL error for the parameters, $\alpha_{\perp}$, $\alpha_{\parallel}$, $f \sigma_8$ from different datasets.}
\begin{tabular}{ccccc}
\hline\hline
Samples&$\alpha_{\perp}$ &$\alpha_{\parallel}$ &$f \sigma_8$ & $\chi^2/{\rm dof}$  \\ \hline  
ELG & $ 0.921 \pm 0.077 $	&	$ 1.083 \pm 0.128 $	&	$0.304 \pm 0.081 $ & $ 167/138$  \\ 
ELG, SGC & $0.959\pm 0.089 $	&	$1.107 \pm 0.142  $	&	$ 0.332 \pm 0.113 $ & $90/67$  \\
ELG &$ \rm fixed $	&	$ \rm fixed $	&$0.402 \pm 0.041 $ & $ 170/140$  \\ \hline
LRGpCMASS &	$1.016 \pm 0.021$	& $1.007 \pm 0.028$&	$0.472 \pm 0.043$  &  $161/138$   \\
LRGpCMASS &	$\rm fixed$	& $\rm fixed $&	$0.448 \pm 0.032$  &  $161/140$   \\ \hline
CROSS &$0.949 \pm 0.040 $	&	$1.118 \pm 0.118 $	&	$0.342 \pm 0.085 $ &$147/138 $   \\ 
CROSS &$ \rm fixed $	&	$\rm fixed $	&	$0.443 \pm 0.050 $ &$148/140 $   \\ \hline 
ELG+LRGpCMASS & $1.000 \pm 0.020 $	&	$1.021 \pm 0.027 $	&$0.419 \pm 0.037 $  & $308/279$    \\
ELG,SGC +LRGpCMASS  & $1.012 \pm 0.022  $	& $1.015 \pm 0.029  $	&$0.442 \pm 0.042  $ &$228/208 $  \\
ELG+CROSS & $0.960 \pm 0.037$	&	$ 1.048 \pm 0.074  $	&	$ 0.380 \pm 0.063 $&$286/279$  \\
CROSS+LRGpCMASS &$1.006 \pm 0.021 $	&	$1.016 \pm 0.029 $	&	$0.444 \pm 0.041 $ & $298/279$  \\ \hline
Joint ($N_p=10$) & $ 1.000 \pm 0.020 $	&$ 1.014 \pm 0.029  $	&$ 0.432 \pm 0.038  $ & $ 410/422$   \\
Joint ($N_p=10, \rm w/\, AP\, fixed$) &$\rm fixed $	&	$ \rm fixed $	& $0.440 \pm 0.028$ & $408 /424$   \\
Joint ($N_p=12$) & $1.001 \pm 0.020 $	&	$1.016 \pm 0.029 $	&$0.432 \pm 0.038 $ & $ 412/420$   \\
Joint ($N_p=12, \rm w/\, AP\, fixed$)  & $\rm fixed $	&	$ \rm fixed $	& $0.442 \pm 0.029  $  & $ 410 /422$   \\
\hline\hline                      
\end{tabular}
\label{tab:data_fit}
\end{table*}

\section{Cosmological implications}
\label{sec:application}

In this section, we briefly discuss cosmological implications of our joint measurements from the multi-tracer analysis. 

We use the distance measurements to constrain the geometry of the Universe in the framework of a non-flat $\Lambda$CDM cosmology, in which the Hubble expansion rate is 
\ba 
H(z) = H_0 \sqrt{\Omega_{m0}(1+z)^3+\Omega_{\Lambda 0} + (1-\Omega_{m0}-\Omega_{\Lambda 0})}\,.
\ea
To avoid the dependence on $r_d$, we work in the parameter space of $(\Omega_{m0}, \Omega_{\Lambda 0}, H_0r_d)$. 

The BAO datasets used here include the isotropic BAO measurements using MGS \citep{Ross:2014qpa} and 6dFGRS \citep{6dFBAO} galaxy samples; BOSS DR12 anisotropic BAO measurements in the low- and middle-redshift bins, \ie\,$(0.2<z<0.5)$ and $(0.4<z<0.6)$ \citep{Alam2016}; the anisotropic BAO measurement from eBOSS DR16 quasars \citep{Neveux2020, Hou2020}, Lyman-$\alpha$ forest \citep{Bourboux2020}, and our multi-tracer analysis of eBOSS DR16 ELG and LRGpCMASS. 

In Figure \ref{fig:contour_OMOL}, we present the 68 and 95\% CL contour plots (black) for the cosmological parameters $(\Omega_{m0}, \Omega_{\Lambda 0}, H_0r_d)$, and their one-dimensional probability distributions. The joint BAO data sets a strong constraint on dark energy density, $i.e.\,\Omega_{\Lambda0}=0.751 \pm 0.066$. The BAO alone favors the existence of dark energy at the significance of $11 \,\sigma$. Compared with the constraining result ($i.e.$ \, blue contours in Figure \ref{fig:contour_OMOL}) \citep{Ata:2017dya} using the isotropic BAO measurements using MGS \citep{Ross:2014qpa} and 6dFGRS \citep{6dFBAO} galaxy samples; the anisotropic BAO measurement in three $z$ bins from BOSS DR12 \citep{Alam2016}; the isotropic BAO measurement from eBOSS DR14 quasars \citep{Ata:2017dya}; and BOSS DR11 and DR12 Lyman-$\alpha$ sample \citep{Font2013wce, Bautista:2017zgn}, the significance of non-zero dark energy density is improved by a factor of $1.67$. 

\section{Conclusions} 
\label{sec:conclusion}
We perform a multi-tracer analysis in configuration space using the final eBOSS LRG sample combined with the BOSS CMASS sample, and the final eBOSS ELG sample.   

We test the validity of the multi-tracer pipeline using the $N-$body MDPL2 mocks and EZmocks, before applying to the analysis of real data. We report a high-precision measurement on the cosmic expansion rate and growth of structure at the effective redshift $z=0.77$, and find an improvement in the FoM of the $\alpha_{\perp}$, $\alpha_{||}$, $f\sigma_8$ parameters of $11\%$ over that using the LRGpCMASS sample alone. Note that the area covered by the LRGpCMASS sample is larger by a factor of $13$ than that of the ELG sample, thus the LRGpCMASS dominates the information content in the joint analysis. Even in this case, a non-trivial improvement in the FoM is contributed by the ELG sample, demonstrating the efficacy of the multi-tracer method.

We combine our measurement with previous BAO distance measurements from MGS, 6dFGS, BOSS DR12, and new BAO distance measurements from eBOSS DR16 quasars and eBOSS DR16 Lyman-$\alpha$ sample, to test a non-flat $\Lambda$CDM cosmology. It is found that a non-zero dark energy density is favored by BAO alone at a $11 \,\sigma$ significance.

The stage-IV galaxy surveys, such as the Dark Energy Spectroscopic Instrument (DESI),\footnote{\url{https://www.desi.lbl.gov/}} and Euclid,\footnote{\url{https://www.euclid-ec.org/}} aim to observe multiple tracers with high density at higher redshifts. These surveys will explore the history of cosmic expansion and growth of structure with higher precision, taking advantage of the multi-tracer nature of the survey. Admittedly, this requires a concerted effort to minimize systematics, both through better theoretical modeling and a deeper understanding of observational effects.

\section*{Data Availability}
The correlation functions, covariance matrices, and resulting likelihoods for cosmological parameters are available via the SDSS Science Archive Server (\url{https://svn.sdss.org/public/data/eboss/mcmc/trunk/}), and also available at \url{https://github.com/ytcosmo/MultiTracerBAORSD/}.

\section*{Acknowledgements}
GBZ is supported by the National Key Basic Research and Development Program of China (No. 2018YFA0404503). YW and GBZ are supported by NSFC Grants 11890691, 11925303, 11720101004 and 11673025. YW is also supported by the Nebula Talents Program of NAOC. GBZ is also supported by a grant of CAS Interdisciplinary Innovation Team. SA and JAP are supported by the European Research Council through the COSFORM Research Grant (\#670193). OHEP acknowledges funding from the WFIRST program through NNG26PJ30C and NNN12AA01C. GR acknowledges support from the National Research Foundation of Korea (NRF) through Grants No. 2017R1E1A1A01077508 and No. 2020R1A2C1005655 funded by the Korean Ministry of Education, Science and Technology (MoEST), and from the faculty research fund of Sejong University.

Funding for the Sloan Digital Sky Survey IV has been provided by the Alfred P. Sloan Foundation, the U.S. Department of Energy Office of Science, and the Participating Institutions. SDSS-IV acknowledges
support and resources from the Center for High-Performance Computing at
the University of Utah. The SDSS web site is \url{http://www.sdss.org/}. 

SDSS-IV is managed by the Astrophysical Research Consortium for the 
Participating Institutions of the SDSS Collaboration including the 
Brazilian Participation Group, the Carnegie Institution for Science, 
Carnegie Mellon University, the Chilean Participation Group, the French Participation Group, Harvard-Smithsonian Center for Astrophysics, 
Instituto de Astrof\'isica de Canarias, The Johns Hopkins University, Kavli Institute for the Physics and Mathematics of the Universe (IPMU) / 
University of Tokyo, the Korean Participation Group, Lawrence Berkeley National Laboratory, 
Leibniz Institut f\"ur Astrophysik Potsdam (AIP),  
Max-Planck-Institut f\"ur Astronomie (MPIA Heidelberg), 
Max-Planck-Institut f\"ur Astrophysik (MPA Garching), 
Max-Planck-Institut f\"ur Extraterrestrische Physik (MPE), 
National Astronomical Observatories of China, New Mexico State University, 
New York University, University of Notre Dame, 
Observat\'ario Nacional / MCTI, The Ohio State University, 
Pennsylvania State University, Shanghai Astronomical Observatory, 
United Kingdom Participation Group,
Universidad Nacional Aut\'onoma de M\'exico, University of Arizona, 
University of Colorado Boulder, University of Oxford, University of Portsmouth, 
University of Utah, University of Virginia, University of Washington, University of Wisconsin, 
Vanderbilt University, and Yale University.

This work made use of the facilities and staff of the UK Sciama High Performance Computing cluster supported by the ICG, SEPNet and the University of Portsmouth. This research used resources of the National Energy Research Scientific Computing Center, a DOE Office of Science User Facility supported by the Office of Science of the U.S. Department of Energy under Contract No. DE-AC02-05CH11231. The authors are pleased to acknowledge that the work reported in this paper was substantially performed using the Princeton Research Computing resources at Princeton University which is consortium of groups including the Princeton Institute for Computational Science and Engineering and the Princeton University Office of Information Technology's Research Computing department.

\bibliographystyle{mn2e}
\setlength{\bibhang}{2.0em}
\setlength\labelwidth{0.0em}
\bibliography{multitracer}

\label{lastpage}

\end{document}